%% file: article_revised.tex
\documentclass[10pt,english,notitlepage,a4paper,superscriptaddress,floatfix,longbibliography]{revtex4-2}

\usepackage[T1]{fontenc}
\usepackage{amsmath}
\usepackage{amssymb}
\usepackage{graphicx}
\usepackage{bm}
\usepackage{ulem}
\usepackage[unicode=true,pdfusetitle,
urlcolor=blue,bookmarks=true,bookmarksnumbered=false,bookmarksopen=false,
 breaklinks=false,pdfborder={0 0 1},backref=false,colorlinks=true]
 {hyperref}
\usepackage{comment}
\usepackage[percent]{overpic}
\usepackage{url}
\makeatletter

\newcommand{\im}{\mathrm{i}}

\usepackage{xcolor}
\usepackage{tikz}
\definecolor{myblue}{RGB}{12, 12, 158}
\definecolor{myred}{RGB}{158, 19, 22}
\definecolor{myorange}{RGB}{245, 150, 12}
\definecolor{mygreen}{RGB}{26, 148, 49}
\definecolor{Prune}{RGB}{99,0,60}
\definecolor{Purple}{RGB}{75, 0, 130}
\definecolor{Pink}{RGB}{255, 105, 180}
\definecolor{deepskyblue}{RGB}{0, 191,255}
\definecolor{limegreen}{RGB}{50, 205, 50}
\definecolor{crimson}{rgb}{0.86, 0.08, 0.24}
\definecolor{cerulean}{rgb}{0.0, 0.48, 0.65}
\definecolor{electricpurple}{rgb}{0.75, 0.0, 1.0}

\definecolor{blue(ncs)}{rgb}{0.0, 0.53, 0.74}

\newcommand{\figpanel}[1]{(\textbf{\lowercase{#1}})}

\makeatother

\begin{document}
\title{The Symmetric Perceptron: a Teacher-Student Scenario}

\author{Giovanni Catania}
\affiliation{Departamento de Física Teórica, Universidad Complutense de Madrid,
28040 Madrid, Spain}
\affiliation{Institute for Cross-disciplinary Physics and Complex Systems IFISC (CSIC-UIB),
Campus Universitat Illes Balears, 07122 Palma de Mallorca, Spain.}
\author{Aur\'{e}lien Decelle}
\affiliation{Departamento de Física Teórica, Universidad Complutense de Madrid,
28040 Madrid, Spain}
\affiliation{Université Paris-Saclay, CNRS, INRIA Tau team, LISN, 91190 Gif-sur-Yvette,
France.}
\affiliation{GISC - Grupo Interdisciplinar de Sistemas Complejos 28040 Madrid, Spain.}
\author{Suhanee Korpe}
\email{suhanee21@iiserb.ac.in}
\affiliation{Departamento de Física Teórica, Universidad Complutense de Madrid,
28040 Madrid, Spain}
\affiliation{Indian Institute of Science Education and Research, Bhopal 462066, India}

\begin{abstract}
\noindent We introduce and solve a teacher-student formulation of the symmetric binary Perceptron, turning a traditionally storage-oriented model into a planted inference problem with a guaranteed solution at any sample density. We adapt the formulation of the symmetric Perceptron which traditionally considers either the u-shaped potential or the rectangular one, by including labels in both regions. With this formulation, we analyze both the Bayes-optimal regime at for noise-less examples and the effect of thermal noise under two different potential/classification rules. Using annealed and quenched free-entropy calculations in the high-dimensional limit, we map the phase diagram in the three control parameters, namely the sample density $\alpha$, the distance between the origin and one of the symmetric hyperplanes $\kappa$ and temperature $T$, and identify a robust scenario where learning is organized by a second-order instability that creates teacher-correlated suboptimal states, followed by a first-order transition to full alignment. 
We show how this structure depends on the choice of potential, the interplay between metastability of the suboptimal solution and its melting towards the planted configuration, which is relevant for Monte Carlo-based optimization algorithms.

\end{abstract}

\maketitle

\section{Introduction}

\noindent The Perceptron~\cite{rosenblatt1958perceptron} is the simplest supervised classification model and plays a dual role: it constitutes a foundational element of modern deep-learning architectures~\cite{lecun2015deep} and provides a paradigmatic framework for analyzing learning dynamics~\cite{engel2001statistical,coolen2005theory} and applying statistical-mechanics techniques to learning problems~\cite{zdeborova2016statistical}. Since Gardner’s seminal work on the storage capacity of the spherical Perceptron~\cite{gardner1988space}, and the work by Krauth \& Mezard on its binary version~\cite{krauth_storage_1989} a large body of research has explored its thermodynamic properties, phase transitions, and algorithmic behavior using tools originally developed for disordered systems~\cite{mezard_parisi_virasoro,sompolinskylearningfromexamples,baldassi_efficient_2007,baldassi_generalization_2009}. In particular, the teacher-student (TS) setting~\cite{sompolinsky1990learning,zdeborova2016statistical} has emerged as a natural framework to study learning as an inference problem, where a student model attempts to recover a hidden teacher~\cite{achlioptas2008algorithmic,krzakala2009hiding} from labeled examples generated according to the same rule~\cite{statmechlearningruleBiel,baldassi_unreasonable_2016}.

In this work, we focus on a \textit{symmetric} version of the binary Perceptron in a TS setting, where the solution space is symmetric under the inversion of the weight vector. Unlike the standard Perceptron, whose decision boundary is a single hyperplane, the symmetric Perceptron is invariant under a global sign flip of the weights, leading to a classification rule defined by two symmetric hyperplanes, the distance of which is tuned by the so-called margin parameter $\kappa$. This symmetry introduces a richer structure in the space of solutions and naturally gives rise to multiple competing phases. Symmetric versions of the Perceptron have previously been studied mainly in the context of storage problems with prescribed label distributions, often using u-shaped or rectangular potentials~\cite{Aubin_2019,Barbier_2024}. Here, we adapt this framework to the TS scenario by explicitly incorporating teacher-generated labels in both regions defined by the symmetric decision rule. To our knowledge, an analytic mean-field characterization of this model in the TS setting, including the full phase-diagram structure --- in terms of the three control parameters, namely the sample density $\alpha$, the margin parameter $\kappa$ and temperature $T$ --- has not been reported previously. 
As we show, this discrete symmetry qualitatively changes the learning landscape: it can separate the onset of non-zero but sub-optimal correlations towards the teacher from full teacher recovery, producing a characteristic second-order/first-order sequence of transitions.

Our analysis combines annealed and quenched calculations of the free entropy in the high-dimensional limit, using the replica method under a replica-symmetric ansatz~\cite{mezard_parisi_virasoro}. We investigate both a piecewise constant loss, analogous to the original formulation studied by Gardner, and a linear loss that penalizes misclassified samples proportionally to their distance from the decision boundary~\cite{engel2001statistical}. This allows us to characterize not only the Bayes-Optimal (BO), zero-temperature regime—where the student perfectly matches the teacher—but also the effects of thermal noise and finite temperature learning dynamics.

\noindent Our goal is to characterize the phase diagram of the symmetric binary Perceptron as a function of the sample density $\alpha$, the margin $\kappa$, and the temperature $T$. To this end, we compute the relevant transition lines — first- and second-order transitions as well as spinodal points of the sub-optimal/optimal solution — that determine when information about the teacher becomes accessible and when full recovery is thermodynamically favored. We also identify a phase characterized by metastable, suboptimal states, in which the system admits solutions that are uncorrelated or partially correlated with the teacher. These transitions control the onset of learning, the development of correlations with the teacher, and the stability of paramagnetic and glassy phases. By systematically comparing different potentials and temperatures, we obtain a comprehensive thermodynamic picture of learning in symmetric Perceptron models.

This paper is organized as follows, in Section II, we introduce the symmetric binary Perceptron model under the TS setting and the two potentials, namely the constant and the linear potentials, under which we study the model. Next, in Section III, we begin our analysis by presenting the annealed computation for the constant potential. We provide the expressions for the free energy and the corresponding saddle-point equations. We also present the free-energy profiles that clearly depict the phase transitions, spinodal, first-order, and second-order, for different values of $\alpha$ and 
$\kappa$ at $T=0$, and summarize the resulting phase behavior. To obtain a more rigorous description, we then present the quenched computations and results in Section IV. In particular, we analyze the Bayes-optimal case and consider both potentials at finite temperature. We show the free-energy profiles for the Bayes-optimal case by varying $\alpha$ and 
$\kappa$ and compare the ($T$, $\alpha$) phase diagrams for the constant and linear potentials. Finally, we conclude by summarizing our results and discussing possible extensions and future directions of this work. We provide detailed calculations in the appendices. In Appendix A, we present the derivation of the annealed disorder computation. In Appendix B, we provide the quenched analysis, with three subsections deriving the results for the Bayes-optimal case, the constant potential, and the linear potential, respectively.

\section{Definition of the model}

\begin{figure}
    \centering
\begin{overpic}[width=.4\textwidth]{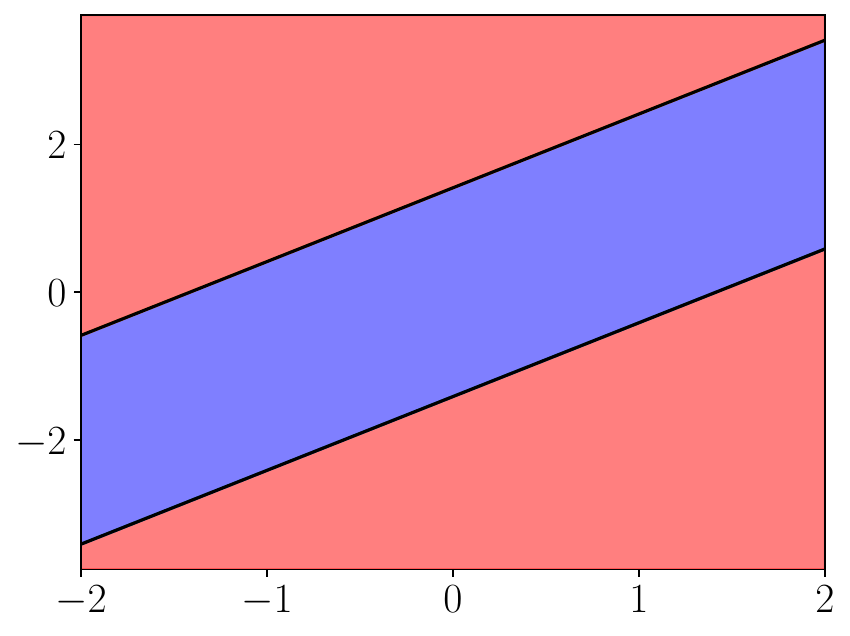}
\put(20,55.5){{$-1$}}
\put(50,37.5){{$1$}}
\put(80,15.5){{$-1$}}
\end{overpic}
    \caption{Decision boundary for a symmetric Perceptron in $N=2$ with $w=\left[-1 \slash \sqrt{2}, 1 \slash \sqrt{2}\right]$ and $\kappa=1$, illustrating a case where the data in the red region are labelled $+1$, while the ones in the blue region are labeled $-1$.}
    \label{fig:hyperplanes}
\end{figure}

\noindent The classical binary Perceptron formulation consists of classifying a set of $M$ binary examples $\bm{\xi}^\mu$, where each sample lives in a $N-$dimensional hypercube $\bm{\xi}^\mu \in  \{-1,1 \}^N $. In our analysis, we consider the high-dimensional regime where the number of samples $M$ scales with the system size $N$, such that $\alpha = M \slash N \sim O(1)$ as  $M, N \rightarrow \infty$.  The output of a Perceptron is typically considered to be generated by the (noise-less) learning rule $\sigma^{\mu} = {\rm sign}(\bm{w} \cdot \bm{\xi}^{\mu} )$, where $\sigma_\mu$ is the output that takes binary values $
\{-1,1\}$. Geometrically, this implies that a single hyperplane orthogonal to $\bm{w}$ separates the data points into two classes. In the context of a TS scenario, the labels of the samples are generated by a teacher Perceptron with a weight vector $\bm{w}_0$,  whose components are assumed to be i.i.d. binary variables $\bm{w}_{0 i}\in \{-1,1\}$. The student Perceptron learns its weights $\bm{w}$ using the training sample $\bm{\xi}^\mu$ and its corresponding label $\sigma^\mu_0$ generated by the teacher, according to
\begin{equation}
    \sigma_0^{\mu} = {\rm sign}(\bm{w}_0 \cdot \bm{\xi}^{\mu} ) .\label{eq:signRuleTeacher}
\end{equation}
In practice, the label is determined by the datum's position relative to the separating hyperplane defined by the teacher. In this paper, we consider a symmetric version of the Perceptron where if $\bm{w}$ is a solution to our problem, that is, it classifies all samples correctly, then so is $\bm{w}' = -\bm{w}$. This  choice symmetrizes the model, so that the configuration space is now cut by two hyperplanes, defined by $\kappa + \bm{w} \cdot \bm{x} = 0$ and $\kappa - \bm{w} \cdot \bm{x} = 0$, where $\kappa$ represents the shift of the hyperplane from the origin. In such a case, the Hamiltonian will be symmetric under the transformation $\bm{w} \to -\bm{w}$. A simple visualization of the decision boundary in the symmetric Perceptron for a 2-dimensional model ($N=2$) is given in Fig.~\ref{fig:hyperplanes}.

In the storage formulation, the fundamental question that is addressed is usually about the volume of samples that can be classified correctly, given a (typically random) set of $M$ examples, from which the critical capacity is computed and corresponds to a SAT-UNSAT transition in the jargon of constraint satisfaction~\cite{Aubin_2019, Barbier_2024, mezard_information_2009}. A typical choice for theoretical computation is to use random samples and random labels that can be contained inside or outside the area delimited by both hyperplanes. 

In the TS formulation~\cite{sompolinsky1990learning,sompolinskylearningfromexamples}, we have at our disposal a teacher that generates the labels, thereby ensuring that a solution to the problem exists, regardless of the number of samples. In the TS formulation, and in the absence of external thermal noise (i.e., at $T=0$), the teacher guaranties satisfiability for any $\alpha$, so the central question becomes when the Gibbs measure acquires non-trivial overlap with the teacher and how metastability can obstruct algorithmic recovery.

\begin{figure}[t!]
    \centering
    \includegraphics[width=0.42\textwidth]{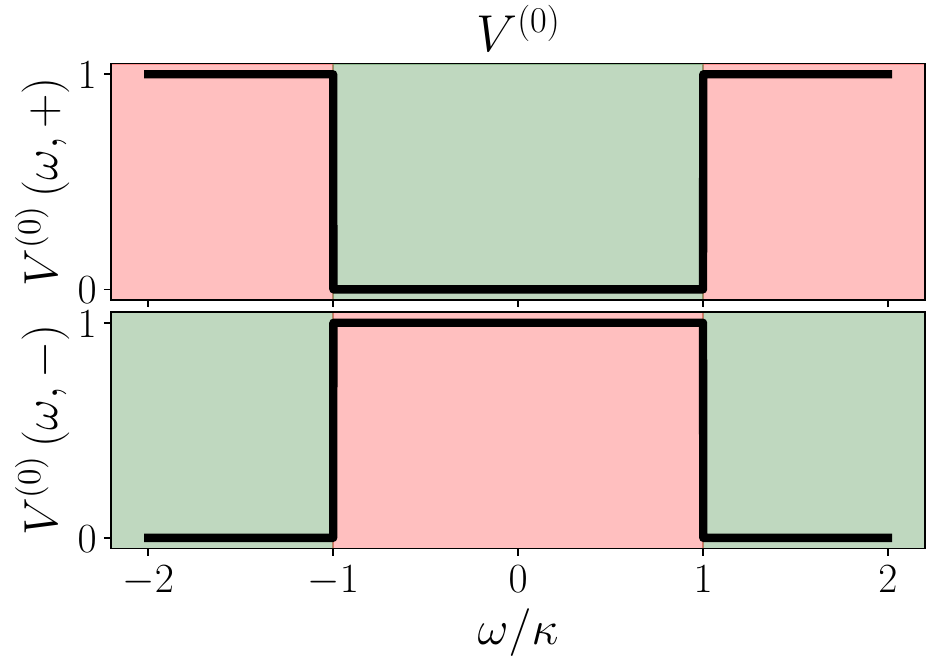}
    \hfill
    \includegraphics[width=0.42\textwidth]{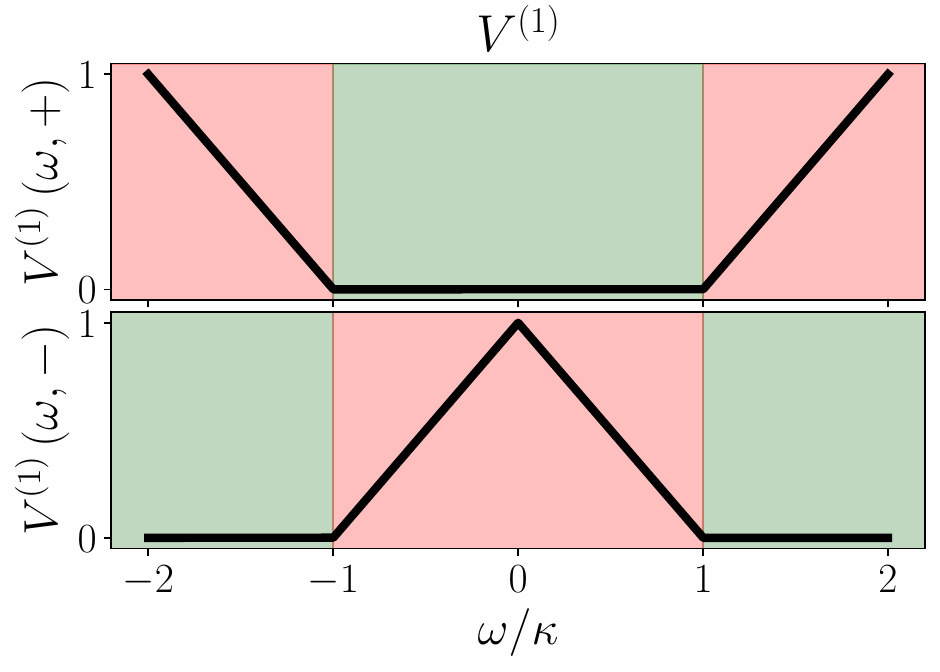}
    \caption{\label{fig:costfunction} \textbf{Left:} the potential $V^{(0)}$ for both labels $\sigma_0 = \pm 1$. With this potential, any error has a fixed cost. \textbf{Right:} the linear potential $V^{(1)}$ for both labels $\sigma_0 = \pm 1$. This potential tends to penalize more errors that are far away from the decision boundary.}
\end{figure}

Let us now formalize the symmetric Perceptron model. We can first define a loss function that will, given the labels of the system and the weights $\bm{w}$, count how many samples are misclassified. The Hamiltonian, or training loss function, associated with the student can be written in the following form
\begin{equation}
    \mathcal{H}[\bm{w}|\underline{\bm{\xi}},\bm{w}_0] = \sum_{\mu=1}^M V(\omega_\mu, \sigma_0^{\mu}),
\end{equation}
\noindent where we used the notation $\omega_\mu = \bm{\xi}^\mu \cdot \bm{w}$ where $\underline{\bm{\xi}}$ denotes the set of patterns $\bm{\xi}^\mu$, $\mu=1,\dots,M$ and $\bm{w}_0$ the teacher's weight with $\sigma_0^\mu$ given by the decision boundary of the teacher's weights as in Fig.~\ref{fig:costfunction}. In practice, if the label of a sample is $\sigma_0 = 1$, it should lie inside the hyperplanes defined by the teacher, while in the case $\sigma_0 = -1$, it should lie outside, as illustrated in Fig.~\ref{fig:costfunction}. When focusing only on correctly classifying all the labels, it is enough to just count the number of errors. In a statistical physics' context, it is also common to consider the finite temperature behaviour, by defining the Gibbs-Boltzmann distribution $p \propto \exp(-\beta \mathcal{H})$ of the system to analyse how entropic contribution can compete with the minimization of the cost function. To extend our analysis to the finite temperature of such object,  we considered two different definitions of the Hamiltonian that are equivalent in the limit $\beta^{-1} = T \to 0$. First, we will consider the piecewise constant potential which assigns a constant cost to incorrectly classified samples as shown in Fig.~\ref{fig:costfunction} left panel. The analytical formulation of the potential is given by
\begin{align}
    V^{(0)}(\omega, \sigma_0) &= 
      \begin{cases}
        \Theta\left[-\omega - \kappa\right] + \Theta\left[\omega - \kappa\right], & \text{if } \sigma_0 = 1,  \\
        \Theta\left[\kappa + \omega\right]\Theta\left[\kappa - \omega\right], & \text{if } \sigma_0 = -1.
    \end{cases} \label{eq:Vconst}
\end{align}

\noindent In this case, any misclassified sample suffers the same unit error: this is typically referred to as the Gibbs learning rule~\cite{gardner1988space, engel2001statistical}. Despite the known problem with the frozen dynamics of such potential~\cite{horner_dynamics_1992}, we consider it to be much easier to deal with in order to solve numerically the phase diagram over the entire temperature range. 

\noindent The second Hamiltonian we consider contains a linear potential which assigns a linear cost to the incorrectly classified data points as shown in Fig.~\ref{fig:costfunction} right panel, similarly to the one studied in~\cite{sompolinskylearningfromexamples}, which leads to a smoother energy landscape. The form of the linear potential is given below
\begin{align}
    V^{(1)}(\omega, \sigma_0) &= 
      \begin{cases}
        \left(-\omega -\kappa\right) \Theta\left[-\omega-\kappa\right]+\left(\omega-\kappa\right)\Theta\left[\omega-\kappa\right], & \text{if } \sigma_0 = 1,  \\
        \left(\kappa - \left| \omega \right| \right)\Theta\left[\kappa+\omega\right]\Theta\left[\kappa-\omega\right], & \text{if } \sigma_0 = -1.
    \end{cases} \label{eq:Vlinear}
\end{align}
This potential follows the traditional Perceptron cost function~\cite{engel2001statistical} in which misclassified samples that are far away from the decision boundary are penalized more strongly. This potential is more suitable for Monte Carlo based optimization, given its informative potential, it is also expected, that when using the constant piece-wise potential $V^{(0)}$, the dynamics will be frozen due to entropic barriers, while in the case of linear potential, it is expected that for a large number of data, any local dynamics could find easily a solution~\cite{Horner1992_DMFT,horner_dynamics_1992}. This comparison isolates which features of the phase diagram are intrinsic to the symmetric decision rule and which depend on the smoothness of the optimization landscape induced by the choice of the loss.

\noindent In order to characterize the phase diagram of the symmetric binary Perceptron, we use the replica approach to implement the mean field theory, \cite{mezard_parisi_virasoro,gardner1988space,charbonneau2023spin} to compute the partition functions and free energies for both the potentials, as a function of the inverse temperature $\beta$, the number of samples $\alpha$ and the width $\kappa$ of the potential
\begin{equation}
    \mathcal{Z}[\underline{\bm{\xi}},\bm{w}_0] = \sum_{\bm{w} = \{\pm 1\}^N} \exp\left(-\beta \mathcal{H}[\bm{w}|\underline{\bm{\xi}},\bm{w}_0] \right). ~\label{eq:part_fun}
\end{equation}where the student's weights are summed over all possible values. In the rest of the paper, we will consider that the components $\xi_i^\mu$ of the dataset and the teacher weights $w_{0i}$ will be distributed uniformly in in $\{\pm 1\}$ with equal probability $p=1/2$, and denote the average over it as $\mathbb{E}_{\mathbb{D}}[.]$ for the dataset and $\mathbb{E}_{\bm{w}_0}[.]$ for the teacher's weight vector.
\section{Annealed Free Energy Computation for the Piece-Wise Potential }

\noindent In order to understand the physics of the model, we provide an analysis of the annealed computation of the free entropy for the Hamiltonian with the piece-wise constant potential. In the annealed approximation, instead of computing the disorder-average of the logarithm of the partition function, we take the logarithm of the average of the partition function. Already at $T=0$, the annealed theory reveals a coexistence structure in the overlap between equilibrium configurations and the teacher's weights that explains why teacher recovery can be discontinuous and is preceded by an extensive regime in $\alpha$ where suboptimal minima dominate the thermodynamics.

In the following, we will denote the disorder averaged on the dataset as $\mathbb{E}_{\mathbb{D}}$ and the average over the teacher weights as $\mathbb{E}_{\bm{w}_0}$. In the thermodynamic limit, the expression of the free energy, denoted by $\mathcal{G}$ is given by
\begin{equation}
    -\mathcal{G}(\kappa,\alpha)_{\rm annealed} =  \lim_{N\to \infty} \frac{1}{N} \log \mathbb{E}_{\mathbb{D},\bm{w}_0} \left[\mathcal{Z}[\underline{\bm{\xi}},\bm{w}_0]\right],
\end{equation}
where the dataset is $\mathbb{D} = \{\bm{\xi}^\mu\}_{\mu=1}^M$, and we analyze the free energy as a function of the margin $\kappa$, the inverse temperature $\beta$ and the ratio of the number of samples and features $\alpha$. We put details of the computation in Appendix~\ref{app:AnnealedFreeEn}. To explain the computation in brief, we introduce the order parameter 
\begin{align}
    R &= \frac{1}{N}\sum_i^N w_i w_{0i} 
\end{align}
representing the overlap between the student and the teacher. A value $R\sim0$ means that there is no correlation between the teacher and the student while $R=1$ means the student has found the teacher configuration. When introducing this parameter, the free energy density as decomposes as given below

\begin{align}
 -\mathcal{G} &= - R \hat{R} + G_S(\hat{R}) + \alpha\,G_E(R)
\end{align}

\noindent where $\hat{R}$ is the conjugate overlap parameter, $G_S$ is the entropic contribution, counting the volume of weight configurations at fixed overlap, and $G_E$ is the energetic contribution, which is the average training cost per example at a fixed student-teacher overlap, representing how well a student aligned with the teacher satisfies the constraints. Learning is governed by a trade-off between entropy, which favors many compatible weight configurations, and energy, which penalizes configurations that poorly satisfy the training constraints. While opening the Hamiltonian $V\left(\Delta_{\mu}\right) $ in the energetic part, we split the $\Delta$-integral into disjoint regions according to whether $|\Delta|>\kappa$ or $|\Delta|\le\kappa$, yielding separate Gaussian contributions from the inactive and active regions, respectively, the latter acquiring an additional Boltzmann weight $e^{-\beta}$. Evaluating these pieces explicitly leads to the decomposition of the free energy integral.

\noindent We obtain the free energy of the system by considering the saddle point of the free energy parameterized by the order parameter $R$ and its conjugate parameter $\hat{R}$. The saddle-point solution reflects a competition between the entropic term and the energetic term. The precise derivation of the annealed free energy is given in the Appendix~\ref{app:AnnealedFreeEn}. In order to find the equilibrium of the model in the thermodynamic limit, we determine the minima of this free energy by solving numerically the self-consistent saddle point equations (\eqref{eq:Ranneal} and \eqref{eq:RhatselfAnnealed} given in Appendix~\ref{app:AnnealedFreeEn}). In this approximation, the expression for the free energy is given by
\begin{widetext}
\begin{align*}
& \mathcal{- G} = -R\hat{R} + \ln2\cosh{\hat{R}} + \alpha  \ln \left[2 \int_{\kappa}^{\infty} Du \left[ \Phi \left( \frac{\kappa - uR}{ \sqrt{1 - R^2}} \right) + \Phi \left( \frac{\kappa + uR}{ \sqrt{1 - R^2}} \right) \right] \right. \\
& \left. + \int_{-\kappa}^{\kappa} \frac{Du}{2} \left[ \operatorname{erf} \left( \frac{\kappa + uR}{\sqrt{2} \sqrt{1 - R^2}} \right) + \operatorname{erf} \left( \frac{\kappa - uR}{\sqrt{2} \sqrt{1 - R^2}} \right) \right] + 2 \int_{-\kappa}^{\kappa} Du \left[ \Phi \left( \frac{\kappa - uR}{ \sqrt{1 - R^2}} \right) + \Phi \left( \frac{\kappa + uR}{\sqrt{1 - R^2}} \right) \right]  e^{-\beta}  \right],
\end{align*}
\end{widetext}
where $Du = \exp(-u^2 / 2)/\sqrt{2\pi}du$ and $\Phi(x) = {\rm erfc}(x/\sqrt{2})/2$. We expect that due to the symmetry of the model, the annealed computation exhibits both a kind of paramagnetic phase where no signal can be detected in the equilibrium measure and thus $R = 0$, and a phase where, when enough samples are provided the system is capable of retrieving at least partially the teacher $R \geq 0$. In such scenario, we can investigate analytically the stability of the paramagnetic solution, and we found that it is stable up to
\begin{align}
\alpha_c^{(2)} = \frac{\pi \left( \text{erfc}^2(\frac{\kappa}{\sqrt{2}}) 
 +\text{erf}^2(\frac{\kappa}{\sqrt{2}}) + 2 \text{erf}(\frac{\kappa}{\sqrt{2}}) \text{erfc}(\frac{\kappa}{\sqrt{2}}) e^{-\beta} \right)}{4 e^{-\kappa^2} \kappa^2 (1- e^{-\beta})}, \label{eq:2ndorder_annealed}
\end{align} 
where a second order phase transition would take place. However, the model exhibits a richer phenomenology due to a phase coexistence between the aforementioned paramagnetic state or a suboptimal solution with $0<R<1$ and the teacher configuration $R=1$. The key qualitative point is that the free-energy landscape can develop three competing minima (paramagnetic, suboptimal correlated, and teacher), whose crossings generate first-order lines and whose disappearances define spinodal points. 

\begin{figure}[t!]
\centering
\begin{overpic}[width=\textwidth]{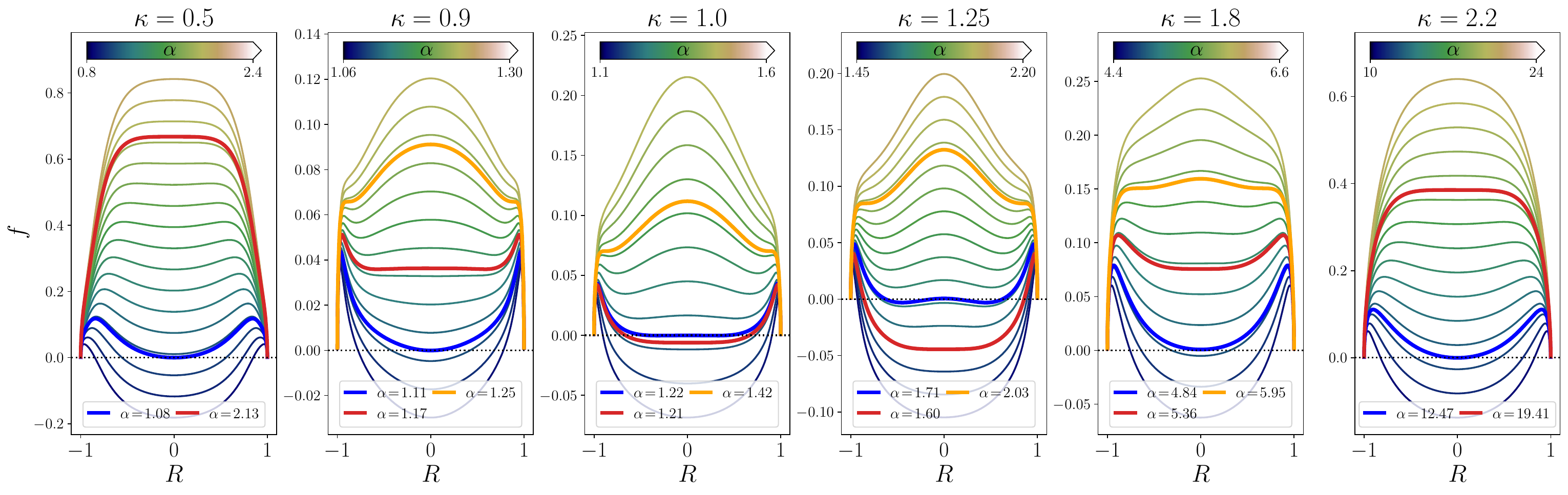}
\put(5,30.5){{\figpanel{a}}}
\put(21,30.5){{\figpanel{b}}}
\put(37.6,30.5){{\figpanel{c}}}
\put(54,30.5){{\figpanel{d}}}
\put(70,30.5){{\figpanel{e}}}
\put(86,30.5){{\figpanel{f}}}
\end{overpic}
\caption{Snapshots of free energy profiles at $6$ different values of $\kappa$, corresponding to the thin vertical lines in Fig.\ref{fig:PD_bothAnnealed_andQuenched} (left panel). Each panel \figpanel{a}$\to$\figpanel{f} shows several free energy profiles $f(R)$ at different values of $\alpha$. The thicker curves correspond to the values of the first/second order transition and the spinodal (when different). In blue we illustrate first order transitions, in red second order ones and in yellow the melting (spinodal) point of the sub-optimal solution $0<R<1$. The colors used here are the same as in Fig. \ref{fig:PD_bothAnnealed_andQuenched}. \label{fig:Free_energies_Annealed_T0}}
\end{figure}
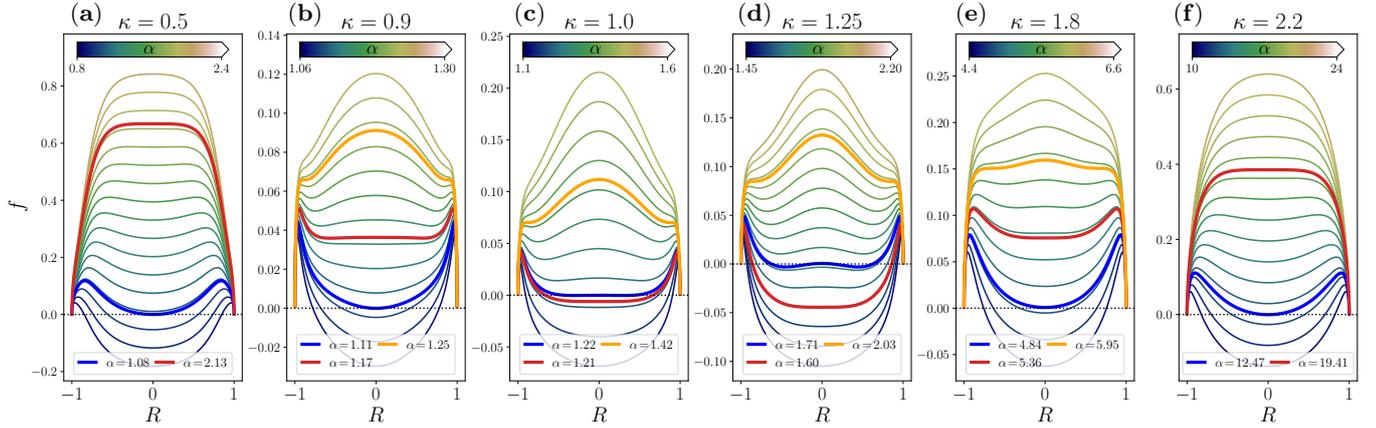
To have a clearer understanding, we summarize in Fig.~\ref{fig:Free_energies_Annealed_T0} six different free-energy profiles as a function of $R$ at different values of $k$ (one profile per each panel) and varying the fraction of samples $\alpha$ available to the students (different values of $\alpha$ are shown in each panel where $\alpha$ increases from blue-ish to yellow-ish colors). For small values of $\kappa$, (see e.g the left-most panel $\kappa=0.5$) and varying the density of samples $\alpha$, the system exhibits a first order phase transition at $\alpha_c^{(1)}$  ($\alpha_c^{(1)} \approx 1.08$ in Fig.~\ref{fig:Free_energies_Annealed_T0}-\figpanel{a}) from a paramagnetic regime $R=0$, towards a system fully polarized towards the teacher, i.e $R=1$. This crossing identifies the first-order transition: at $\alpha_c^{(1)}$ the global minimum jumps from $R=0$ to $R=1$ while the paramagnetic state remains locally stable up to its spinodal point.  
Of course, the teacher is in practice unreachable unless the system is initialized close to it in that regime: indeed,  the paramagnetic solution remains stable for quite large values of $\alpha \sim 2.13$ (given by Eq.~\ref{eq:2ndorder_annealed}) until it melts toward the teacher. Therefore, this second threshold of $\alpha$ (red line in Fig~\eqref{fig:Free_energies_Annealed_T0}) marks the spinodal point of the paramagnetic solution. Increasing $\kappa$ modifies the phenomenology of the model. For a value of $\kappa=0.9$ (depicted in Fig.~\ref{fig:Free_energies_Annealed_T0}-\figpanel{b}), after undergoing the first order transition at $\alpha_c^{(1)} \sim 1.11$, the locally stable paramagnetic solution is split, undergoing a second order phase transition (thick red line in Fig.~\ref{fig:Free_energies_Annealed_T0}), into two (symmetric) suboptimal solutions but with a large overlap with the teacher $R>0$. In this case at $\alpha_c^{(2)} \sim 1.17$, the second order phase transition now takes place within the subdominant paramagnetic. At higher values of $\alpha$, the two suboptimal solutions finally melt toward the teacher at $\alpha \sim 1.25$ (thick orange line). In another intermediate regime at higher values of $\kappa$, e.g in panels \figpanel{c}-\figpanel{d}, the second order phase transition occurs before the first order one. In panel~\figpanel{e} the phenomenology is the same as in ~\figpanel{b}. Finally, increasing further $\kappa$ restores the initial behavior: e.g. at $\kappa=2.2$ (panel~\figpanel{f}), the situation goes back to the extremely small-$\kappa$ case, with a first order transition taking place beforehand and no suboptimal solution with $0<R<1$ exists at any value of $\alpha$. \\

\begin{figure}[t!]
    \centering    \includegraphics[width=\textwidth]{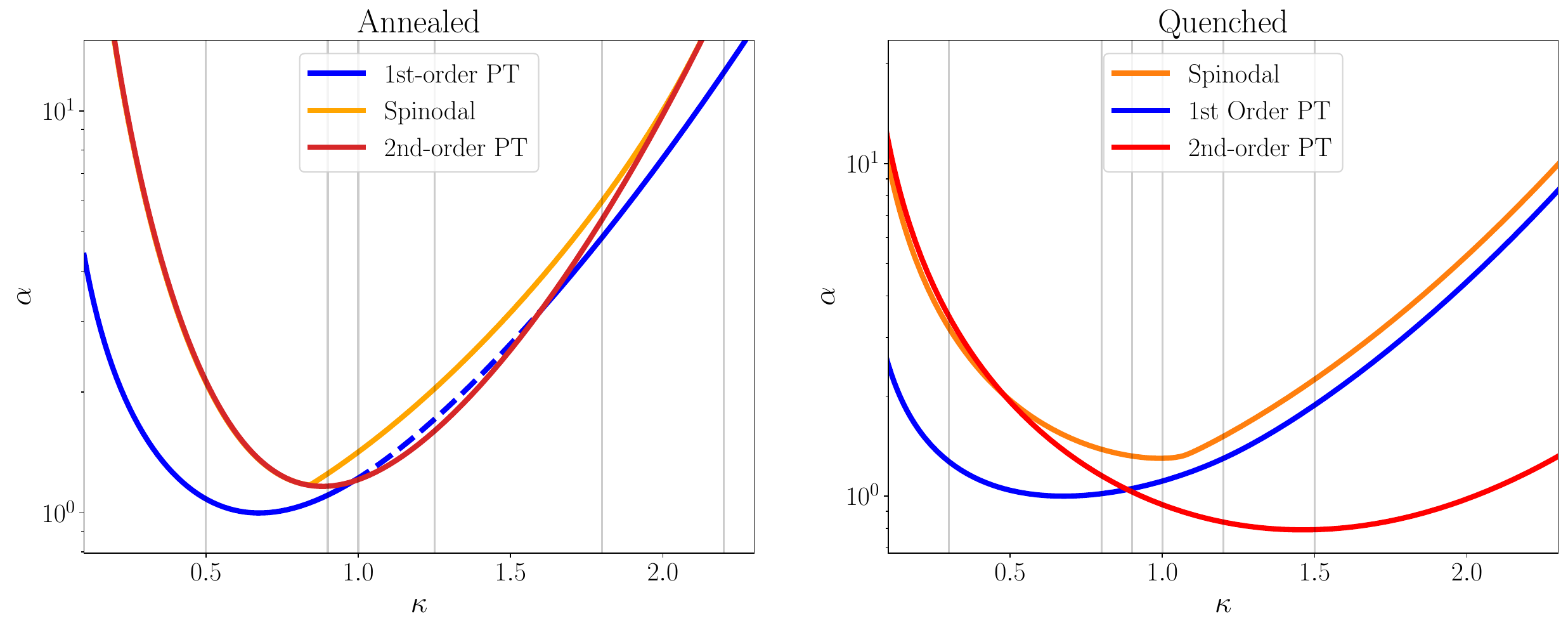}
    \caption{\textbf{Left:} Phase diagram of the symmetric binary Perceptron at $T=0$, using the annealed approximation. \textbf{Right:}  Phase diagram of the symmetric binary Perceptron at $T=0$, using the quenched approximation in the Bayes-Optimal setting. The vertical lines indicate the values of $\kappa$ used on Fig.~\ref{fig:Free_energies_Annealed_T0} (resp.~Fig.~\ref{fig:Free_energies_Quenched_T0XXX}) for the annealed (resp. quenched) case.
    In both cases, the blue line indicated the first order transition, the red line the second order one and the yellow line the spinodal, that is when the sub-optimal solution disappear. In the range of $\kappa$ where the second order phase transition occurs, the dashed-blue line only corresponding to when the unstable paramagnetic state and the teacher state have the same free energy.}
    \label{fig:PD_bothAnnealed_andQuenched}
\end{figure}

\noindent We summarize these different phases in Fig.~\ref{fig:PD_bothAnnealed_andQuenched} (left panel), where we show the various critical lines in the $\left(\kappa-\alpha\right)$ plane at temperature $T=0$. The 3-phases regime with a sub-optimal solution $0<R<1$ exists for intermediate values of $\kappa$, whose spinodal point is depicted in Fig.~\ref{fig:PD_bothAnnealed_andQuenched} as an orange line. From a numerical point of view, the $1-$st order phase transition (blue lines) is computed for a fixed $\kappa$ and varying $\alpha$ by looking at the point where the free energy of the sub-optimal solution (or the paramagnetic one) becomes positive (i.e larger than the teacher's one which is null). Similarly, the spinodal line corresponds to the point at which starting from an initial condition $R \approx 0$ the saddle point equation converges to the teacher $R = 1$. The code used to generate the critical lines in Fig.~\ref{fig:Free_energies_Annealed_T0} (and the other phase diagrams shown in the next section) is included in a open-access repository~\cite{repogithub}.

\section{QUENCHED FREE ENERGY COMPUTATION}

We now turn our analysis towards the quenched free energy for the model defined by Eq.~\eqref{eq:part_fun}.
The main thermodynamic quantity providing information about the system is the quenched average  of the free energy density, given by \begin{equation}
-\mathcal{G}(\kappa,\alpha, \beta) =  \lim_{N\to \infty} \frac{1}{N} \mathbb{E}_{\mathbb{D},\bm{w}_0}\log \mathcal{Z}.
\end{equation} 
The quenched computation is essential in the TS setting because it determines which teacher-correlated states dominate in the \textit{typical} case, beyond the optimistic annealed picture.  To calculate the expectation of the logarithm of the partition function, we use the replica trick, which involves introducing $n$ replicas and using the following identity 
\begin{equation}
\langle \log \mathcal{Z}\rangle   = \lim_{n \to 0} \frac{\langle \mathcal{Z}^n \rangle - 1}{n}
\end{equation}

\noindent By using the replica trick, we end up defining the usual order parameters
\begin{align}
    R_a &= \frac{1}{N}\sum_i^N w_i^a w_{0i} \\
    q_{ab} &= \frac{1}{N}\sum_i^N w_i^a w_i^b ,
\end{align}
representing respectively the overlap between the student and the teacher and the overlap between different students. The first parameter $R_a$ has the same interpretation as in the annealed case, while the second parameter $q_{ab}$ is the overlap between two replicas $a$ and $b$, indicating the presence of potentially glassy states. In general, as in the annealed case, the parameter $R_a$ indicates the overlap with the teacher, and $R_a \sim 1$ means that the system has found the optimal solution. In such a case, we also have that $q_{ab} \sim 1$. A glassy state is detected by both a null overlap with the teacher $R_a \sim 0$, while the overlap between two students is different from zero $q_{ab} \neq 0$.  In order to find the equilibrium of the model in the thermodynamic limit, we assume the replica symmetric (RS) ansatz $R = R_a$ and $q = q_{ab}$ $\forall$ replicas $a$ and $b$, and proceed similarly to the annealed case. In this work, we obtain the free energy of the system by considering the saddle point of the free energy parameterized by the order parameters in the RS approximation. In this section, we first start by analyzing the Bayes-Optimal (BO) results, and finally, we discuss the behavior in temperature of the two different Hamiltonians.

\begin{figure}
\centering
\begin{overpic}[width=\textwidth]{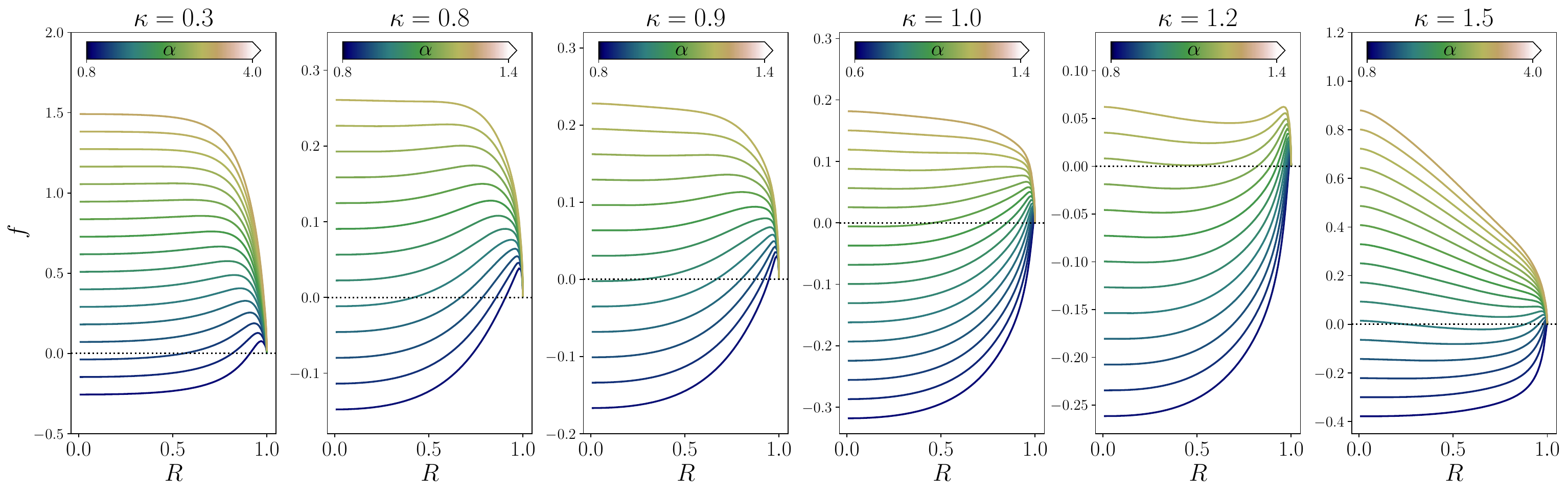}
\put(5,30.5){{\figpanel{a}}}
\put(21,30.5){{\figpanel{b}}}
\put(37.6,30.5){{\figpanel{c}}}
\put(54,30.5){{\figpanel{d}}}
\put(70,30.5){{\figpanel{e}}}
\put(86,30.5){{\figpanel{f}}}
\end{overpic}
\caption{Snapshots of free energy profiles at 6 different values of $\kappa$ at $T=0$ in the Bayes-Optimal setting. We can observe a phase diagram that is qualitatively similar to the annealed case. From small values of $\kappa$, we observe first a first order phase transition, follow by a melting toward the teacher state at higher values of $\alpha$. Then, we have the at larger $\kappa$ a second order phase transition before the melting.
\label{fig:Free_energies_Quenched_T0XXX}}
\end{figure}
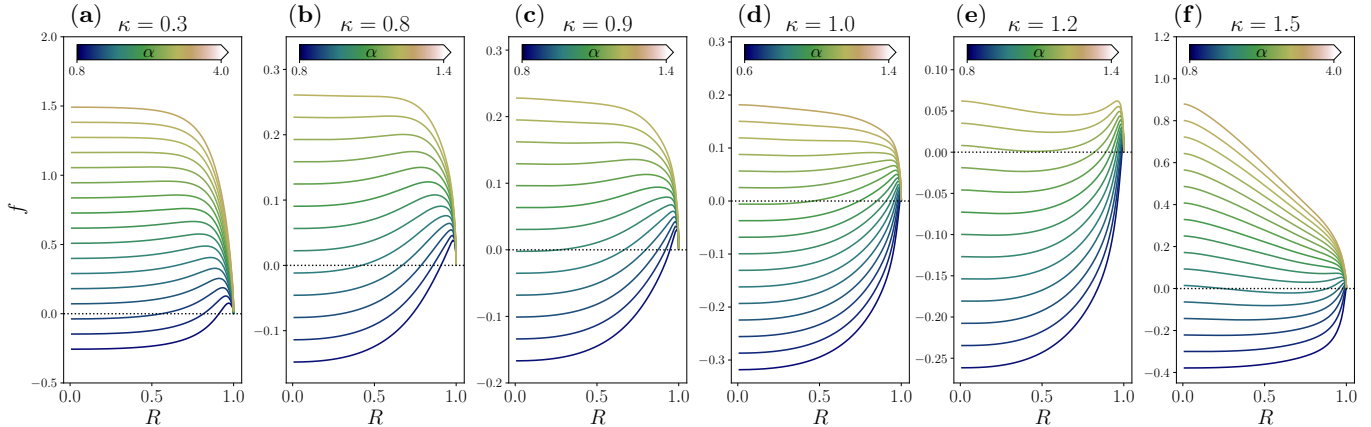

\subsection{The Bayesian-Optimal case}
\noindent The Bayes-Optimal scenario, corresponds to the limit of zero temperature ($T=0$) where the student model perfectly matches the teacher's learning algorithm. As established in the analysis of the classical Perceptron ~\cite{engel2001statistical}, this condition significantly simplifies the computation. The system is constrained to the Nishimori line~\cite{zdeborova2016statistical}, where the overlap between two replicas of the student ($q$) becomes identical to the overlap between the student and the teacher ($R$). Consequently, setting $q = R$ in the replica symmetric ansatz yields the following simplified free energy,
\begin{align*}
\mathcal{- G} & = - \frac{\hat{R}}{2}\left( R + 1\right) + \int Dz\ln2\cosh{(\sqrt{\hat{R}}z + \hat{R})}  + \alpha \times \left[ \int Dt\left[\Phi\left(A_{+}\right)+\Phi\left(A_{-}\right)\right]\log\left\{ \left[\Phi\left(A_{+}\right)+\Phi\left(A_{-}\right)\right]\right\}  \right.\\ 
& \left. +\int Dt\left[\frac{1}{2}\text{erf}\left(\frac{A_{+}}{\sqrt{2}}\right)+\frac{1}{2}\text{erf}\left(\frac{A_{-}^{}}{\sqrt{2}}\right)\right]\log\left\{\frac{1}{2}\text{erf}\left(\frac{A_{+}}{\sqrt{2}}\right)+\frac{1}{2}\text{erf}\left(\frac{A_{-}}{\sqrt{2}}\right)\right\}\right]
\end{align*}
with 
\begin{equation}
 A_{\pm}=\frac{\kappa\pm\sqrt{R}t}{\sqrt{1-R}}
\end{equation}
and $\Phi(x)=\frac{1}{2}\text{erfc}(\frac{x}{\sqrt{2}})$. In this regime, the only order parameter is therefore given by
\begin{align*}
R &= \int Dz \tanh^2 \left( \sqrt{\hat{R}} z + \hat{R} \right) 
\end{align*}
and the expression for the conjugate parameter $\hat{R}$ can be found in Appendix~\hyperref[app:BO]{B2}. We can fully appreciate the different phase to which the system goes through as the parameter $\alpha$ is changed and for various values of $\kappa$. Similar to the annealed case, we can compute the threshold up to which the paramagnetic solution is stable, given by the following formula:
\begin{equation}
    \alpha_c^{(2)} = \frac{\pi}{2 \kappa^2}e^{\kappa^2}  {\rm erf}\left(\kappa/\sqrt{2}\right){\rm erfc}\left(\kappa/\sqrt{2}\right).
\end{equation}
The main difference w.r.t. the annealed case is that now the sub-optimal solution $0<R<1$ seems to always exist for $\kappa$ larger than a threshold identified by the point where the orange and red lines in Fig.~\ref{fig:PD_bothAnnealed_andQuenched} start to have different values.

\subsection{Behavior in temperature}

\noindent We finally exhibit the behavior of the system at finite temperature. In this case, both potentials are expected to exhibit different behaviors. We direct the reader to Appendix \hyperref[app:piecewise_potential]{B1} and \hyperref[app:Vlinear]{B3}  for the detailed computation and only write the functional form of the free energy for both potentials here. For the potential $V^{(0)}$ we obtain 

\begin{widetext}
    \begin{align}
\frac{1}{n}\log G_{E}	&=\int Dt\left[\Phi\left(A_{+}\right)+\Phi\left(A_{-}\right)\right]\log\left\{ 2\left[\Phi\left(B_{+}\right)+\Phi\left(B_{-}\right)\right]+e^{-\beta}\left[\text{erf}\left(\frac{B_{+}}{\sqrt{2}}\right)+\text{erf}\left(\frac{B_{-}}{\sqrt{2}}\right)\right]\right\} + \nonumber\\ 
	&+\int Dt\left[\frac{1}{2}\text{erf}\left(\frac{A_{+}}{\sqrt{2}}\right)+\frac{1}{2}\text{erf}\left(\frac{A_{-}^{}}{\sqrt{2}}\right)\right]\log\left\{ 2e^{-\beta}\left[\Phi\left(B_{+}\right)+\Phi\left(B_{-}\right)\right]+\text{erf}\left(\frac{B_{+}}{\sqrt{2}}\right)+\text{erf}\left(\frac{B_{-}}{\sqrt{2}}\right)\right\}, 
\end{align}
\end{widetext}
with 
\begin{equation}
A_{\pm}=\frac{\kappa\sqrt{q}\pm Rt}{\sqrt{q-R^{2}}}\qquad\text{and}\qquad B_{\pm}=\frac{\kappa\pm\sqrt{q}t}{\sqrt{1-q}}.
\end{equation}
We clearly see how the Boltzmann factor $e^{-\beta}$ contributes to an extra term in the free energy that takes into account the possibility to commit an error penalized by a weight $\beta$. It is possible to compute the instability of the paramagnetic solution at all temperatures and values of $\kappa$, thanks to the spin-flip symmetry of the problem. To do that, we expand the free energy around $R=0$ and look for the point when its second derivative computed in $R=0$ changes sign. For this potential, we find that the instability is given by
\begin{widetext}
\begin{equation}
    \alpha_{\text{2}^{\text{nd}}}^{(0)} \left(\beta,\kappa\right)	= e^{\kappa^2}\frac{\pi \left[ \text{erf}\left(\frac{\kappa}{\sqrt{2}}\right)+e^{-\beta}\text{erfc}\left(\frac{\kappa}{\sqrt{2}}\right) \right] \left[ e^{-\beta}\text{erf}\left(\frac{\kappa}{\sqrt{2}}\right)+\text{erfc}\left(\frac{\kappa}{\sqrt{2}}\right) \right]}{2\kappa^2 \left(1-e^{-2\beta}\right)}\label{eq:alpha_1storder}
\end{equation}
\end{widetext}
from which we recover the case of the BO case in the limit $\beta \to \infty$.
The free energy in the case of the $V^{(1)}$ potential takes the following form

\begin{widetext}
\begin{align}
\frac{1}{n}\log G_{E}	&=\int Dt\left[\Phi\left(A_{+}\right)+\Phi\left(A_{-}\right)\right]\log\left\{ \left[\Phi\left(B_{+}\right)+\Phi\left(B_{-}\right)\right]+e^{-\beta \kappa +\frac{\beta^2 (1-q)}{2}}P_1\right\} + \nonumber\\ 
	&+\int Dt\left[\frac{1}{2}\text{erf}\left(\frac{A_{+}}{\sqrt{2}}\right)+\frac{1}{2}\text{erf}\left(\frac{A_{-}^{}}{\sqrt{2}}\right)\right]\log\left\{ e^{\beta \kappa +\frac{\beta^2 (1-q)}{2}}P_2+\frac{1}{2}\text{erf}\left(\frac{B_{+}}{\sqrt{2}}\right)+\frac{1}{2}\text{erf}\left(\frac{B_{-}}{\sqrt{2}}\right)\right\}, \\
    P_1 &= \frac{1}{2} e^{-\beta \sqrt{q}t}\left[{\rm erf}\left(\frac{B_{-}(\kappa)}{\sqrt{2}}\right) - {\rm erf}\left(\frac{B_{-}(0)}{\sqrt{2}}\right) \right] + \frac{1}{2}e^{\beta \sqrt{q}t}\left[{\rm erf}\left(\frac{B_{+}(0)}{\sqrt{2}}\right) - {\rm erf}\left( \frac{B_{+}(-\kappa)}{\sqrt{2}}\right) \right], \\
    P_2 &=  \frac{1}{2}e^{\beta \sqrt{q}t}{\rm erfc}\left(\frac{B_{+}(\kappa)}{\sqrt{2}}\right) + \frac{1}{2}e^{-\beta \sqrt{q}t}{\rm erfc}\left(\frac{-B_{-}(-\kappa)}{\sqrt{2}}\right).
\end{align}
\end{widetext}
In this expression, the temperature dependent terms are now more complicated, since they have to take into account the distance from the decision boundary. The computation in this case of the instability is more involved, yet it can be analytically done and yields
    \begin{align}
    \frac{1}{\alpha_c^{(1)}} &= e^{-\kappa^2} \kappa \beta \left\{2 e^{\frac{\beta ^2}{2}} \text{erf}\left[\frac{\kappa -\beta }{\sqrt{2}}\right]+2 e^{\frac{\beta ^2}{2}} \text{erf}\left[\frac{\beta
}{\sqrt{2}}\right] \right. \nonumber\\
& \left. +\text{erf}\left[\frac{\kappa}{\sqrt{2}}\right] \left(2 e^{\frac{\kappa^2}{2}}-2 e^{\kappa \beta }+e^{\frac{1}{2} \left(\kappa^2+\beta ^2\right)} \sqrt{2
\pi } \beta  \text{erf}\left[\frac{\kappa-\beta }{\sqrt{2}}\right]+e^{\frac{1}{2} \left(\kappa^2+\beta ^2\right)} \sqrt{2 \pi } \beta  \text{erf}\left[\frac{\beta
}{\sqrt{2}}\right]\right) \right. \nonumber\\
& \left. +2 e^{\kappa \beta } \text{erfc}\left[\frac{\kappa}{\sqrt{2}}\right]+2 e^{\frac{1}{2} (\kappa+\beta )^2} \text{erfc}\left[\frac{\kappa+\beta
}{\sqrt{2}}\right]-2 e^{\frac{1}{2} \beta  (4 \kappa+\beta )} \text{erfc}\left[\frac{\kappa+\beta }{\sqrt{2}}\right]-\right. \nonumber\\ & -\left.e^{\frac{1}{2} \left(\kappa^2+4 \kappa \beta +\beta ^2\right)} \sqrt{2 \pi } \beta  \text{erfc}\left[\frac{\kappa}{\sqrt{2}}\right] \text{erfc}\left[\frac{\kappa+\beta
}{\sqrt{2}}\right]\right\}  \nonumber\\
& \slash \left\{\pi  \left(e^{\frac{\beta ^2}{2}} \text{erf}\left[\frac{\kappa-\beta }{\sqrt{2}}\right]+e^{\frac{\beta ^2}{2}}
\text{erf}\left[\frac{\beta }{\sqrt{2}}\right]+e^{\kappa \beta } \text{erfc}\left[\frac{\kappa}{\sqrt{2}}\right]\right) \left(\text{erf}\left[\frac{\kappa}{\sqrt{2}}\right]+e^{\frac{1}{2}
\beta  (2 \kappa+\beta )} \text{erfc}\left[\frac{\kappa +\beta }{\sqrt{2}}\right]\right)\right\}. \label{eq:alpha_2ndorder_v1}
\end{align}
Taken together, Eqs. (\ref{eq:alpha_1storder}) and (\ref{eq:alpha_2ndorder_v1}), these expressions show that temperature does not merely smear transitions; it reshapes the stability of the $R=0$ phase in a loss-dependent way, thereby controlling whether learning begins continuously or only via a discontinuous jump. 

\begin{figure}[t!]
    \begin{overpic}[width=.45\textwidth]{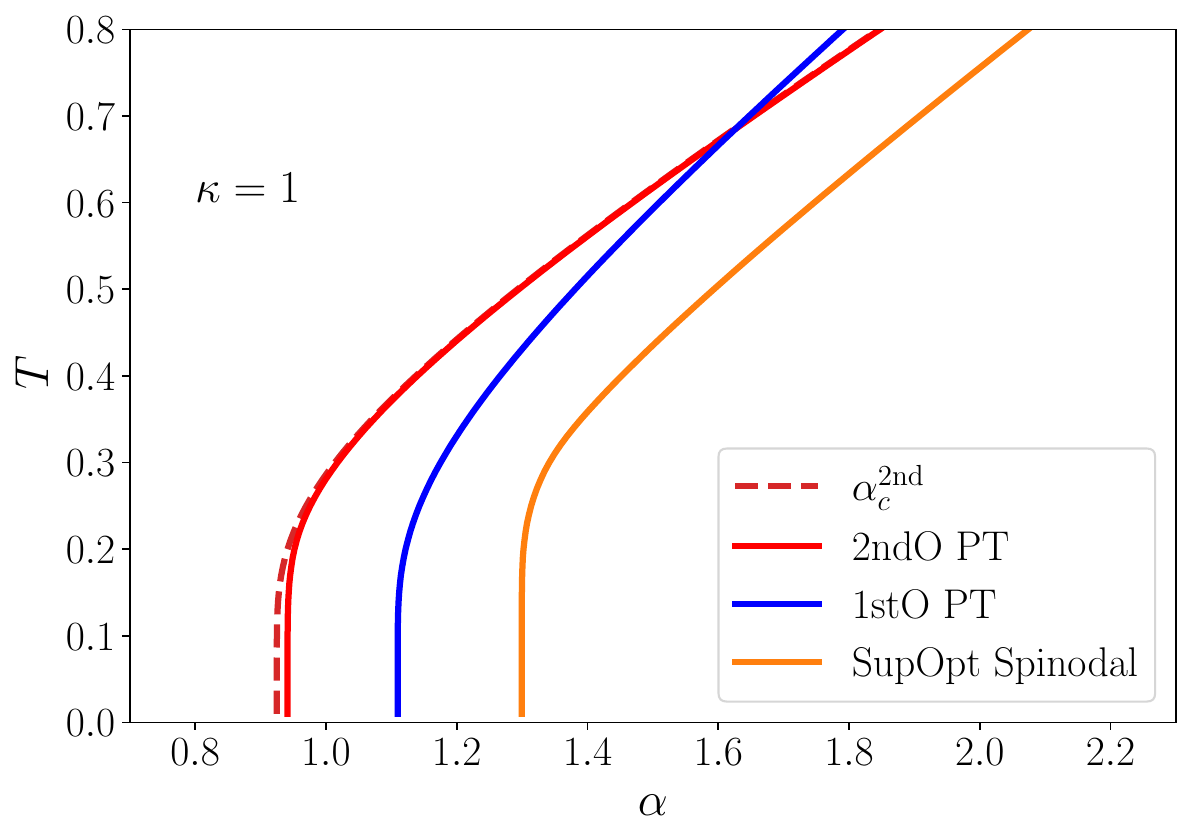}
    \put(13,63){{\figpanel{a}}}
\end{overpic}
    \hfill 
    \begin{overpic}[width=.45\textwidth]{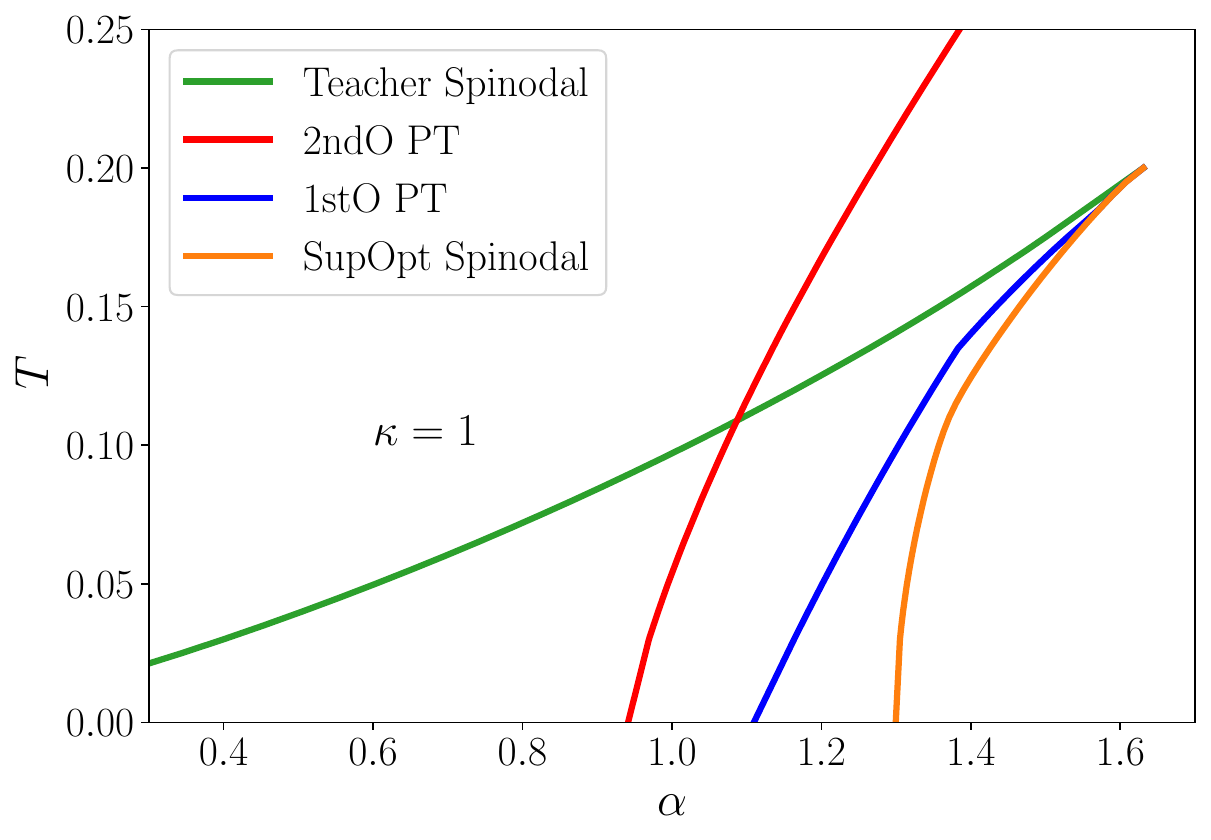}
\put(90,62.5){{\figpanel{b}}}
\end{overpic}
\caption{\textbf{Left:} Phase diagram of the constant potential $V_0$ at $\kappa=1$. \textbf{Right:} Phase diagram of the piece-wise constant potential $V_1$ at $\kappa=1$. As in the case of the Perceptron, the orange line is expected to terminate at $(0,0)$. In this parameter regime, the shape of the potential strongly influences the structure of the loss landscape. For $V_0$, and for sufficiently large $T$ and $\alpha$, there exists a phase in which the teacher dominates the equilibrium measure, while a typical experiment remains trapped in completely uncorrelated states. In contrast, for $V_1$, the second-order phase transition is always dominant with respect to the other transitions and provides information about the teacher well before the latter even becomes a metastable state.}
    \label{fig:PD_temperature}
\end{figure}

\noindent In the following, we provide the full phase diagram in the $\alpha-T$ plane of the system for the specific case $\kappa=1$. On Fig.~\ref{fig:PD_temperature} (left), we plot the different lines of transitions. We can see that for $T<0.7$, by increasing $\alpha$ we gradually pass from a paramagnetic regime $q=R=0$ where the teacher does not exist toward first, the teacher appearing (teacher spinodal line), then a second order phase transition where the paramagnetic state splits into two sub-optimal states, then a first order transition where the teacher state becomes dominant to end up with the melting of the sub-optimal states. For high temperature, the first and second order transitions are inverted as can be seen. At the dynamical level, it is expected that the system is frozen, as already seen for the piece-wise potential of the usual Perceptron~\cite{horner_dynamics_1992}. On Fig.~\ref{fig:PD_temperature} (right), we plot the same kind of figure when considering the potential $V_1$. In this phase diagram, we qualitatively recover the part of the same physics as the classical Perceptron. The main difference now is that a second order transition makes its appearance, and thus the system, for sufficiently high temperatures, remains trapped in a paramagnetic state $R=0$ until it crosses the second order phase transition.

\section{Conclusions}
\noindent In this paper, we have presented a statistical-mechanical analysis of the symmetric binary Perceptron in a TS scenario. By extending the traditional formulation of the symmetric Perceptron to explicitly include teacher-generated labels, we were able to study learning as an inference problem rather than a pure storage task. This distinction ensures the existence of a solution for all sample densities and shifts the focus to the nature of convergence toward the teacher as the amount of data (here tuned by the parameter $\alpha$ in the thermodynamic limit) increases. Our central finding is that the sign symmetry generically splits learning into two stages: the \textit{appearance} of teacher correlation via a second-order instability and the \textit{selection} of the teacher via a first-order transition, with a metastable regime in between.

Using both annealed and quenched free entropy calculations, we characterized the phase structure of the model in the thermodynamic limit. In the Bayes-optimal, zero-temperature regime, we found that learning can proceed through a sequence of phase transitions whose order depends sensitively on the margin $\kappa$. In particular, the system may exhibit a second-order transition from a paramagnetic phase to suboptimal states with finite overlap with the teacher, followed by a first-order transition toward full alignment. This behavior highlights the nontrivial role played by symmetry and margin constraints in shaping the learning landscape.

At finite temperature, the phenomenology becomes even richer. For the piecewise constant potential, thermal fluctuations introduce additional metastability and spinodal lines, reminiscent of the frozen dynamics observed in earlier studies of the Perceptron~\cite{horner_dynamics_1992}. In contrast, the linear potential leads to smoother energy landscapes and phase diagrams that more closely resemble those of the classical Perceptron, while still retaining signatures of symmetry-induced transitions absent in the standard model. Our results show that the choice of loss function has a qualitative impact on both the thermodynamic phases and the expected dynamical behavior of learning algorithms. This thermodynamic organization provides a direct mechanism for algorithmic variability: local dynamics can remain trapped in suboptimal teacher-correlated states up to spinodals, while smoother losses can reduce barriers without removing the underlying symmetry-driven transition structure.

\noindent While our analysis is performed within the replica symmetric approximation, one can already anticipate that, in the $T=0$ regime, the easy–hard algorithmic threshold is connected to instabilities of the RS solution, namely to the de Almeida–Thouless line~\cite{de1978stability}, in close analogy with the (classical) Perceptron case. Indeed, at $T=0$ computations are carried out in the Bayes-optimal regime, and the emergence of suboptimal solutions is typically associated with the failure of message-passing algorithms to converge, which in turn signals the onset of the AT instability.

\noindent Overall, this work provides an integrated perspective on the interplay between symmetry in the decision boundary, margin constraints, and noise in high-dimensional learning problems. Although we focus on a toy model, the symmetric Perceptron admits a fully analytic mean-field description within the framework of spin-glass theory, making it a valuable theoretical laboratory. Beyond its conceptual interest, this model serves as a minimal setting to investigate learning scenarios characterized by inherent degeneracies and competing solutions.

\noindent Several directions for future research naturally emerge. On the theoretical side, it would be important to address replica-symmetry-breaking effects at low temperatures, as well as to investigate the behavior of the AT line~\cite{de1978stability} in temperature and to characterize the emergence of metastable states and their impact on the performance of Monte Carlo-based optimization algorithms, as recently emphasized in~\cite{angelini_limits_2023,catania2023copycat}. In addition, within the TS framework developed here, it would be interesting to study the behavior of the model in the presence of interacting copies, where many similar systems are coupled together~\cite{baldassi_unreasonable_2016,catania2023copycat,angelini_limits_2023,angelini2025algorithmic}. In particular, one could investigate how the second-order phase transition identified in this work affects the properties of the robust ensemble, both in the presence and absence of quiet planting. In the Perceptron setting~\cite{catania2023copycat}, it has been shown in the TS regime that coupling multiple copies of the same system tends to suppress the dynamical transition toward glassy phases, leading to a marked improvement in recovering the teacher. A natural extension of the present work would therefore be to understand whether similar mechanisms operate in the symmetric Perceptron, and how they influence both the algorithmic landscape and the generalization properties of the model.

\section*{Acknowledgements}
\noindent Authors acknowledge financial support by the Comunidad de Madrid and the Complutense University of Madrid through the Atracción de Talento program (Refs. 2019-T1/TIC-13298 \& Refs. 2023-5A/TIC-28934), the project PID2021-125506NA-I00 financed by the ``Ministerio de Economía y Competitividad, Agencia Estatal de Investigación" (MICIU/AEI/10.13039/501100011033), the Fondo Europeo de Desarrollo Regional (FEDER, UE).

\bibliographystyle{apsrev4-2}
\bibliography{biblio}

\appendix

\setcounter{figure}{0}  
\renewcommand{\thefigure}{S\arabic{figure}}

\input{appendix_in}

\end{document}

%% file: appendix_in.tex
\renewcommand{\thesubsection}{\thesection \arabic{subsection}}
\counterwithin*{equation}{section}
\counterwithin*{equation}{subsection}

\section{Annealed Free Energy\label{app:AnnealedFreeEn}}
\noindent In this section of the appendix, we provide the calculations for the computation of the annealed free energy for the piecewise potential defined in Eq.~\eqref{eq:Vconst}. According to the standard spin-glass calculation of the annealed disorder, we take the log of the averaged partition function. Here we use the indices \(\mu \in \{1, \ldots, M\}\) to denote the labeled data points and \(i \in \{1, \ldots, N\}\) to denote the components of the weight vector. We write the partition function as 
\begin{equation}
\mathcal{Z} = \exp \left[ 
    -\beta \sum_{\mu}^M V^0\left(\Delta_{\mu}\right) 
\right].
\end{equation}

\noindent We start by defining the stabilities as 

\begin{equation}
\Delta_\mu = \sigma_\mu^0 \frac{ \bm{\xi}^\mu \cdot \bm{w}}{\sqrt{N}}, \label{eq:Delta_mu} 
\end{equation}

\noindent where $\sigma_\mu^0$ is the label generated by the teacher Perceptron given by Eq.~\eqref{eq:signRuleTeacher}, $\bm{\xi}^\mu$ is the $N-$dimensional binary examples and $\bm{w}$ is the weight vector. Here, $\bm{w}_0$ is the weight vector corresponding to the teacher Perceptron. Using \eqref{eq:Delta_mu} and \eqref{eq:signRuleTeacher} and applying delta functions and their Fourier representation, the partition function can be written as 
\begin{align}
\mathcal{Z} = &\int \prod_{\mu}^M \frac{d\Delta_\mu d\hat{\Delta}_\mu}{2\pi} \int \prod_{\mu}^M \frac{dw_{0 \mu} d\hat{w}_{0 \mu}}{2\pi} \exp \left[ -\beta \sum_{\mu}^M V^0\left(\Delta_{\mu}\right) \right] \nonumber \\
&\quad \times \exp \left[ \im \sum_{\mu}^M \Delta_\mu \hat{\Delta}_\mu + \im \sum_{\mu}^M w_{0 \mu} \hat{w}_{0 \mu} - \im \sum_{\mu}^M \hat{\Delta}_\mu \sigma_0^\mu \frac{\bm{w} \cdot \bm{\xi}^\mu}{\sqrt{N}} - \im \sum_{\mu}^M \hat{w}_{0 \mu} \frac{\bm{w}_0 \cdot \bm{\xi}^\mu}{\sqrt{N}} \right] \label{eq:Zanneal_after_fourier},
\end{align}
where \(\hat{\omega}_{0 \mu}\) and \(\hat{\Delta}_\mu\) are the conjugate variables of \(\omega_{0 \mu}\) and \(\Delta_\mu\) respectively introduced through a Fourier transform. Now, we average over the components of the weight vector. Here, we assume that the components are i.i.d.with binary entries such that \(\xi_i^\mu \in \{-1, 1\}\). The average over the teacher weight vector, whose components are binary i.i.d  variables $\bm{w}_{0 i}\in \{-1,1\}$, will become trivial in the final expression. Taking the average of the components \(\xi_i^\mu\) in \eqref{eq:Zanneal_after_fourier}, we get 
\begin{align}
& \left\langle 
\exp \left ( -\frac{\im}{\sqrt{N}} \sum_\mu^M \hat{\omega}_{0 \mu} (\bm{w}_0 \cdot \bm{\xi}^\mu) 
- \frac{\im}{\sqrt{N}} \sum_{\mu}^M \hat{\Delta}_{\mu} \sigma_0^\mu (\bm{w}_\mu \cdot \bm{\xi}^\mu) \right )
\right\rangle_{\{\bm{\xi}^\mu\}_{\mu=1}^M} \notag \\
&= \prod_{i,\mu} \left\langle 
\exp \left[ -\frac{\im}{\sqrt{N}} \xi_i^\mu \left( \hat{\omega}_{0 \mu} w_{i 0} + \sigma_0^\mu  \hat{\Delta}_{\mu} w_{i} \right) \right] 
\right\rangle_{\xi_i^\mu} \notag \\
&= \prod_{i,\mu} 2 \cosh \left[ 
\frac{\im}{\sqrt{N}} \left( \hat{\omega}_{0 \mu} w_{i 0} + \sigma_0^\mu \hat{\Delta}_{\mu} w_{i} \right)
\right] \notag \\
&\approx \exp \left[ 
-\frac{1}{2N} \sum_{i,\mu}^{N, M} \left( \hat{\omega}_{0 \mu} w_{i 0} + \sigma_0^\mu \hat{\Delta}_{\mu} w_{i} \right)^2 
\right],
\label{eq:Zanneal_afteravg}
\end{align}
\noindent where in the first line we use the fact that pattern components are i.i.d., and in the last line we expanded for \(N \to \infty\), keeping only the first order, the other ones being subdominant in the thermodynamic limit. 
\begin{align}
\mathcal{Z} &= \int \prod_{\mu}^M \frac{d\Delta_\mu d\hat{\Delta}_\mu}{2\pi} 
\int \prod_{\mu}^M \frac{dw_{0 \mu} d\hat{w}_{0 \mu}}{2\pi} \exp \left[ -\beta \sum_{\mu}^M V^0\left(\Delta_{\mu}\right) \right] \nonumber \\
&\quad \times \exp \Bigg[
    \im \sum_{\mu}^M \Delta_\mu \hat{\Delta}_\mu 
    + \im \sum_{\mu}^M w_{0 \mu} \hat{w}_{0 \mu} 
    - \frac{1}{2} \sum_{\mu}^M (\hat{w}_{0 \mu})^2 - \frac{1}{2} \sum_{\mu}^M (\hat{\Delta}_\mu)^2 
    - \sum_{\mu}^M \hat{\Delta}_\mu \hat{w}_{0 \mu} 
    \left( \sum_{i}^N \frac{w_i w_{0 i}}{N} \right) 
\Bigg].\label{eq:Zanneal_afterexp}
\end{align}
We now introduce the order parameter, namely the overlap between the student and the teacher given by
\begin{align}
R &= \frac{1}{N} \sum_{i}^N w_{0 i} w_i.
\label{eq:Ranneal}
\end{align}
Introducing Eq.~\eqref{eq:Ranneal} in \eqref{eq:Zanneal_afterexp} as a delta function, using the Fourier representation and using $\hat{R'}$ as the conjugate variable for the overlap R, we get 
\begin{align}
\mathcal{Z} &=\int \frac{dR d\hat{R'}}{2\pi/N} \exp \left[ N \left( \im R \hat{R'} + G_S (\hat{R}) + \alpha G_E (R) \right) \right],
\label{eq:simplifiedZ_anneal}
\end{align}
here $G_S (\hat{R}^{'a})$ is the entropic part as it accounts for the volume of the configurations at the fixed overlap R and $G_E (R)$ is the energetic part of the free energy as it is specific to the Hamiltonian or the cost function. Furthermore, $\alpha$ is the ratio between the number of examples and the number of features, mathematically $\alpha = M/N $. The entropic part is given as 
\begin{align}
G_S(\hat{R'}) &= \ln \sum_{i}^N \left(  - \im\hat{R'} w_i w_{0}\right)
\end{align}
and the energetic part as 
\begin{align}
G_E (R) &= \ln \Bigg( \int \frac{dw_0}{\sqrt{2\pi}} 
\int \frac{d\hat{\Delta}}{2\pi} 
\int \prod_a^n d\Delta \exp \Bigg( 
-\frac{w_0^2}{2} - \frac{1}{2} \big(1 - (R)^2\big) (\hat{\Delta})^2 \notag \\
&\quad + \im \Delta \hat{\Delta} - \im w_0 \hat{\Delta} R \Bigg) \Bigg) \exp \left[ -\beta \, V^{(0)}\left(\Delta\right) \right].
\end{align}
\noindent After considering the sum over the binary weights and using $\hat{R}$ to denote $\hat{R} = -\im\hat{R'}$ in $G_S(\hat{R})$, we get the simplified entropic part  as

\begin{align}
G_S(\hat{R}) &=  \ln 2 \cosh (\hat{R}).
\label{eq:GS_annealed}\end{align}
For the energetic part, the Hamiltonian $V^0\left(\Delta\right)$ is opened which leads to breaking down the $\Delta$-integral into separate regions according to whether $|\Delta|>\kappa$ or $|\Delta|\le\kappa$, finally we get
\begin{align}
G_E (R) = & \ln \left[ 2 \int_{\kappa}^{\infty} Du \left( \Phi \left(X_{+}\right) + \Phi \left(X_{-}\right) \right) \right. + \int_{-\kappa}^{\kappa} \frac{Du}{2} \left( \operatorname{erf} \left( X_{+} \right) + \operatorname{erf} \left( X_{-}\right) \right) \left. + 2 \int_{-\kappa}^{\kappa} Du \left( \Phi \left(X_{+}\right) + \Phi \left( X_{-} \right) \right) \,\, e^{-\beta}\right].\label{eq:GE_annealed}
\end{align}
Here after, we denote $w_o$ by the variable u for simplicity and 
\begin{align}
X_{\pm}=\frac{\kappa \pm uR}{ \sqrt{1 - R^2}}.
\end{align}
Additionally, we use  \(D x = e^{-x^2/2} \, \frac{dx}{\sqrt{2\pi}}\), denoting the standard Gaussian probability measure and $\Phi(x)=\frac{1}{2}\text{erfc}(\frac{x}{\sqrt{2}})$. 

Substituting the simplified $G_S(\hat{R})$ and $G_E (R)$ in the partition function \eqref{eq:simplifiedZ_anneal}, we are left with the integral over the overlap R and conjugate $\hat{R}$. This integral can be calculated using the saddle point approximation as the exponent in the integrand  is linear in N and in the thermodynamic limit, we have $N \rightarrow \infty$. This integral is dominated by the saddle points in R and $\hat{R}$.

In order to write down the saddle point equations for the annealed disorder, we introduce $\Omega$ as 
\begin{equation}
    \Omega = \frac{
\mathrm{e}^{-\frac{(\kappa - R u)^2}{2(1 - R^2)}} 
\left( 
\left( 1 + \mathrm{e}^{-\frac{2 \kappa R u}{1 - R^2}} \right) \kappa R 
- \left( 1 - \mathrm{e}^{-\frac{2 \kappa R u}{1 - R^2}} \right) u 
\right)
}{
\sqrt{2 \pi} \left( 1 - R^2 \right)^{3/2}
}.
\end{equation}
\noindent Then the expressions for R and $\hat{R}$ in terms of $\Omega$ are given by
\begin{align}
    R = \tanh{(\hat{R})} \label{eq:Rselfannealed}
\end{align}
\noindent and 
\begin{align}
    \hat{R} = \alpha \frac{\left[ 2 \int_{\kappa}^{\infty} Du (-\Omega) + \int_{-\kappa}^{\kappa} \frac{Du}{2} ( \Omega) + 2 \int_{-\kappa}^{\kappa} Du (-\Omega) \,\,e^{-\beta} \right]}{G_E(R)} \label{eq:RhatselfAnnealed}.
\end{align}
%\giova{is it true that there is a logarithm at the numerator? it seems weird} 
%\textcolor{blue}{SK: you're right, there is no log}
%\giova{great!} 
\noindent Finally, we obtain the free energy for the annealed disorder as 
\begin{align}
-\mathcal{G} &= - R \hat{R} + \ln 2\cosh{\hat{R}} + \alpha \ln \Bigg[ 
2 \int_{\kappa}^{\infty} Du \left( \Phi \left(X_{+}\right) + \Phi \left(X_{-}\right) \right) \notag \quad + \int_{-\kappa}^{\kappa} \frac{Du}{2} \left( \operatorname{erf} \left( X_{+} \right) + \operatorname{erf} \left( X_{-}\right) \right) \notag \\
&\quad + 2 \int_{-\kappa}^{\kappa} Du \left( \Phi \left(X_{+}\right) + \Phi \left( X_{-} \right) \right)\, e^{-\beta} 
\Bigg].
\end{align}
\section{Quenched Free Energy} 
\noindent In this section, we provide the calculations for the computation of quenched free energy for the following three cases: i) the piece-wise potential in \eqref{eq:Vconst}; ii) the Bayes optimality case for the piece-wise potential in \eqref{eq:Vconst} ; and iii) the linear potential  in \eqref{eq:Vlinear}. According to the standard spin-glass calculation of the quenched disorder, we introduce $n$ replicas and take the limit \(n \to 0\). 
\begin{equation}
\langle \log \mathcal{Z}\rangle   = \lim_{n \to 0} \frac{\langle \mathcal{Z}^n \rangle - 1}{n}.
\end{equation}
Here we use the indices \(a, b \in \{1, \ldots, n\}\) to denote the replicas, \(\mu \in \{1, \ldots, M\}\) to denote the labeled data points and \(i \in \{1, \ldots, N\}\) to denote the components of the weight vector. We write the replicated partition function as 
\begin{equation}
\mathcal{Z}^n = \exp \left[ 
    -\beta \sum_{a,\mu}^{n,M} V\left(\Delta_{\mu}^a\right) 
\right].
\end{equation}
We start by defining the stabilities as 
\begin{equation}
\Delta_\mu^a = \sigma_\mu^0 \frac{ \bm{\xi}^\mu \cdot \bm{w}^a}{\sqrt{N}},\label{eq:stabilities_quenched}
\end{equation}
where $\sigma_\mu^0$ is given by \eqref{eq:signRuleTeacher}. Using Eqs. \eqref{eq:stabilities_quenched} and \eqref{eq:signRuleTeacher} and applying delta functions and their Fourier representation, the partition function can be written as 
\begin{align}
\mathcal{Z}^n = &\int \prod_{a,\mu}^{n,M} \frac{d\Delta^a_\mu d\hat{\Delta}^a_\mu}{2\pi} \int \prod_{\mu}^M \frac{d\omega_{0 \mu} d\hat{\omega}_{0 \mu}}{2\pi} \exp \left[ -\beta \sum_{a,\mu}^{n,M} V\left(\Delta_{\mu}^a\right) \right] \nonumber \\
&\quad \times \exp \left[ \im \sum_{\mu,a}^{n,M} \Delta^a_\mu \hat{\Delta}^a_\mu + \im \sum_{\mu}^M \omega_{0 \mu} \hat{\omega}_{0 \mu} - \im \sum_{a,\mu}^{n,M} \hat{\Delta}^a_\mu \sigma_0^\mu\frac{w^a \cdot \xi^\mu}{\sqrt{N}} - \im \sum_{\mu}^M \hat{\omega}_{0 \mu} \frac{w_0 \cdot \xi^\mu}{\sqrt{N}} \right], \label{eq:Zn_afterFourier}
\end{align}
where \(\hat{\omega}_{0 \mu}\) and \(\hat{\Delta}^a_\mu\) are the conjugate variables of \(\omega_{0 \mu}\) and \(\Delta^a_\mu\) respectively introduced through a Fourier transform. Now, we average over the components of the weight vector. Here, we assume that the components are i.i.d. with binary entries such that \(\xi_i^\mu \in \{-1, 1\}\). The average over the teacher weight vector will become trivial in the final expression. Taking the average of the components \(\xi_i^\mu\) in~\eqref{eq:Zn_afterFourier}, we get 
\begin{align}
& \left\langle 
\exp \left ( -\frac{\im}{\sqrt{N}} \sum_\mu^M \hat{\omega}_{0 \mu} (w_0 \cdot \xi^\mu) 
- \frac{\im}{\sqrt{N}} \sum_{a,\mu}^{n,M} \hat{\Delta}_{\mu}^a \sigma_0^\mu (w_\mu^{a} \cdot \xi^\mu) \right )
\right\rangle_{\{\xi^\mu\}_{\mu=1}^M} \notag \\
&= \prod_{i,\mu}^{N,M} \left\langle 
\exp \left[ -\frac{\im}{\sqrt{N}} \xi_i^\mu \left( \hat{\omega}_{0 \mu} w_{i 0} + \sigma_0^\mu \sum_{a}^n \hat{\Delta}_{\mu}^a w_{i}^a \right) \right] 
\right\rangle_{\xi_i^\mu} \notag \\
&= \prod_{i,\mu}^{N,M} 2 \cosh \left[ 
\frac{\im}{\sqrt{N}} \left( \hat{\omega}_{0 \mu} w_{i 0} + \sigma_0^\mu \sum_{a}^n \hat{\Delta}_{\mu}^a w_{i}^{a} \right)
\right] \notag \\
&\approx \exp \left[ 
-\frac{1}{2N} \sum_{i,\mu}^{N,M} \left( \hat{\omega}_{0 \mu} w_{i 0} + \sigma_0^\mu \sum_{a}^n \hat{\Delta}_{\mu}^a w_{i}^{a} \right)^2 
\right],
\end{align}
where in the first line we use the fact that pattern components are i.i.d., and in the last line we expanded for \(N \to \infty\), keeping only the first order, the other ones being subdominant in the thermodynamic limit
\begin{align}
\mathcal{Z}^n &= \int \prod_{a,\mu}^{n,M} \frac{d\Delta^a_\mu d\hat{\Delta}^a_\mu}{2\pi} 
\int \prod_{\mu}^M \frac{dw_{0 \mu} d\hat{w}_{0 \mu}}{2\pi} \exp \left[ -\beta \sum_{a,\mu}^{n,M} V\left(\Delta_{\mu}^a\right) \right] \nonumber \\
&\quad \times \exp \Bigg( 
    \im \sum_{a,\mu}^{n,M} \Delta^a_\mu \hat{\Delta}^a_\mu 
    + \im \sum_{\mu}^M w_{0 \mu} \hat{w}_{0 \mu} 
    - \frac{1}{2} \sum_{\mu}^M (\hat{u}_\mu)^2 - \frac{1}{2} \sum_{a,b}^n \sum_{\mu}^M \hat{\Delta}^a_\mu \hat{\Delta}^b_\mu 
    \left( \sum_{i}^N \frac{w_i^a w_i^b}{N} \right) 
    - \sum_{a,\mu}^{n,M} \hat{\Delta}^a_\mu \hat{u}_\mu 
    \left( \sum_{i}^N \frac{w_i^a w_{0 i}}{N} \right) \notag \\
&\quad - \frac{1}{2} \sum_{\mu}^M \hat{w}_{0 \mu}^2 
    - \sum_{a,\mu}^{n,M} \hat{\Delta}^a_\mu \hat{u}_\mu 
    \left( \sum_{i}^N \frac{w_i^a w_{0 i}}{N} \right)
\Bigg). \label{eq:Zn_quenched_afterav}
\end{align}
The disorder average results in an effective coupling between replicas \(a, b\) and we introduce a set of order parameters, namely the overlap between student  (in replica \(a\)) with the teacher and the overlap between a student vector from two different replicas a and b, respectively given by
\begin{align}
R^a &= \frac{1}{N} \sum_{i}^N w_{0 i} w_i^a \\
q^{ab} &= \frac{1}{N} \sum_{i}^N w_i^a w_i^b .
\end{align}
Substituting these definitions in the expression ~\eqref{eq:Zn_quenched_afterav}, we get
\begin{align}
\mathcal{Z}^n &= \int \prod_{a,b}^n \frac{dq^{ab} d\hat{q}^{ab}}{2\pi/N} \int \prod_{a}^n \frac{dR^a d\hat{R}^a}{2\pi/N} \exp \left( N \left[ i \sum_{a,b}^n q^{ab} \hat{q}^{ab} + i \sum_{a}^n R^a \hat{R}^a + G_S (\hat{q}^{ab}, \hat{R}^a) + \alpha G_E (q^{ab}, R^a) \right] \right),
\end{align}
here $G_S (\hat{q}^{ab}, \hat{R}^a)$ is the entropic part and $G_E (q^{ab}, R^a)$ is the energetic part of the Hamiltonian. Further, $\alpha$ is the ratio between the number of examples and the number of features, mathematically $\alpha = M/N $. We define 
\begin{align}
G_S(\hat{q}^{ab}, \hat{R}^a) &= \ln \sum_{i}^N \left( i \sum_{a,b}^n \hat{q}^{ab} w_i^a w_i^b - i \sum_{a}^n \hat{R}^a w_i^a \right) \label{eq:GS_beforeansatz}
\end{align}
\noindent and 
\begin{align}
G_E (q^{ab}, R^a) &= \ln \Bigg( \int \frac{dw_0}{\sqrt{2\pi}} 
\int \prod_a^n \frac{d\hat{\Delta}^a}{2\pi} 
\int \prod_a^n d\Delta^a \exp \Bigg( 
-\frac{u^2}{2} 
- \frac{1}{2} \sum_a^n \big(1 - (R^a)^2\big) (\hat{\Delta}^a)^2 \notag \\
&\quad - \frac{1}{2} \sum_{a<b}^n \hat{\Delta}^a \hat{\Delta}^b \big(q^{ab} - R^a R^b\big) 
+ i \sum_a^n \Delta^a \hat{\Delta}^a - iu \sum_a^n \hat{\Delta}^a R^a \Bigg) \Bigg) \exp \left[ -\beta \sum_{a}^n V\left(\Delta^a\right) \right]. \label{eq:GE_beforeansatz}
\end{align}
Now, we introduce the simplest ansatz used in spin glass systems known as the replica symmetry case. We consider all the replicas to be identical and we replace  $\hat{q}^{ab} = -i\hat{q}$ and $\hat{R}^a = i\hat{R}$ in Eqs.~\eqref{eq:GS_beforeansatz}-\eqref{eq:GE_beforeansatz} to get the replica symmetric solution. For the entropic part,  $G_S(\hat{q}, \hat{R})$ similar to the annealed calculation, we take the sum over the binary components of the weights. In addition to this, we also use the Hubbard Stratonovich identity to linearize the quadratic weights $w_i^a w_i^b$ coupled with the overlap $q^{ab}$ and introduce the gaussian variable z. Similarly, in the energetic part $G_E (q, R)$, we use Hubbard Stratonovich identity to linearize the quadratic terms and introduce the gaussian variable t. The expressions for $G_S(\hat{q}, \hat{R})$ and $G_E (q, R)$ are given below
\begin{align}
G_S(\hat{q}, \hat{R}) &= - \frac{n \hat{q}}{2} + n \int Dz \ln 2 \cosh (\sqrt{\hat{q}} z + \hat{R})
\end{align}
and
\begin{align}
G_E (q, R) &= \ln \int Dt \int Dw_0 \int \prod_a^n d\Delta^a \int \prod_a d\hat{\Delta}^a \notag \\
&\quad \exp \left( -\frac{1}{2} (1 - q) \sum_a^n (\hat{\Delta}^a)^2 + i \sum_a^n \hat{\Delta}^a (\Delta^a - w_0R - \sqrt{q - R^2}t) \right) \exp \left( -\beta \sum_{a}^n V\left(\Delta^a\right) \right) \notag \\
&= \ln \int Dt \int Dw_0 \left[\int \frac{d\Delta}{\sqrt{2\pi(1-q)}} \right. \notag \\
&\left. \quad \exp \left( -\frac{1}{2}  \frac{(\Delta - w_0R - \sqrt{q - R^2}t)^2}{1 - q} \right) \exp \left( -\beta V\left(\Delta\right) \right) \right]^n. \label{eq:GE_RS_genericpotential}
\end{align}

\subsection{Piece-Wise Potential\label{app:piecewise_potential}}
\renewcommand\theequation{C\arabic{equation}}

\noindent Now, we will open the Hamiltonian for the case of the piece-wise potential using $V^{(0)}\left(\Delta_{\mu}^a\right)$ defined in Eq.~\eqref{eq:Vconst}, once again we split the $\Delta$-integral into separate regions according to whether $|\Delta|>\kappa$ or $|\Delta|\le\kappa$, and after taking the log of $\mathcal{Z}^n$ and sending $n \rightarrow 0$, we get the quenched free energy for the piece-wise potential as
\begin{align}
& -\mathcal{G} = -R\hat{R} + \frac{\hat{q}}{2}\left( q - 1\right) + \int Dz\ln2\cosh{(\sqrt{\hat{q}}z + \hat{R})} \notag \\
& + \alpha \times \left[ \int Dt\left[\Phi\left(A_{+}\right)+\Phi\left(A_{-}\right)\right]\log \left( 2 + (e^{-\beta} - 1) \left[ \text{erf}\left(\frac{B_{+}}{\sqrt{2}}\right) + \text{erf}\left(\frac{B_{-}}{\sqrt{2}}\right) \right] \right)\right.\notag \\ 
& \left. +\int Dt\left[\frac{1}{2}\text{erf}\left(\frac{A_{+}}{\sqrt{2}}\right)+\frac{1}{2}\text{erf}\left(\frac{A_{-}^{}}{\sqrt{2}}\right)\right] \log \left(  2e^{-\beta} - (e^{-\beta} - 1) \left[ \text{erf}\left(\frac{B_{+}}{\sqrt{2}}\right) + \text{erf}\left(\frac{B_{-}}{\sqrt{2}}\right) \right] \right) \right],
\end{align}
where
\begin{align}
A_{\pm}=\frac{\kappa\sqrt{q}\pm Rt}{\sqrt{q-R^{2}}}\qquad\text{and}\qquad B_{\pm}=\frac{\kappa\pm\sqrt{q}t}{\sqrt{1-q}}.
\end{align}
\noindent We use the saddle point approximation to calculate the integral over the variables R, $\hat{R}$, q and $\hat{q}$.
To write the saddle point equations, we first introduce the variables $\Gamma, \Xi$ and $\Psi$ defined respectively as
\begin{align}
\Gamma = \frac{
e^{- \frac{(\kappa \sqrt{q} + R t)^2}{2 (q - R^2)}} \, R \left( 
  - \left( 
    \left( 1 + e^{\frac{2 \kappa \sqrt{q} R t}{q - R^2}} \right) \kappa R 
    + \left( -1 + e^{\frac{2 \kappa \sqrt{q} R t}{q - R^2}} \right) \sqrt{q} t 
  \right) 
\right)
}{
2 \sqrt{2\pi} \sqrt{q} \, (q - R^2)^{3/2}
},
\end{align}

\begin{align}
\Xi= \frac{
e^{- \frac{(\kappa \sqrt{q} + R t)^2}{2 (q - R^2)}} 
\left( 
  \left( 1 + e^{\frac{2 \kappa \sqrt{q} R t}{q - R^2}} \right) \kappa \sqrt{q} R 
  - \left( -1 + e^{\frac{2 \kappa \sqrt{q} R t}{q - R^2}} \right) q t 
\right)
}{
\sqrt{2\pi} (q - R^2)^{3/2}
},
\end{align}
\noindent and
\begin{align}
\Psi = \frac{
e^{\frac{(\kappa - \sqrt{q} \, t)^2}{2 (q - 1)}} q 
\left[ 
\left( 1 + e^{\frac{2 \kappa \sqrt{q} \, t}{q - 1}} \right) \kappa \sqrt{q} 
+ \left( 1 + e^{\frac{2 \kappa \sqrt{q} \, t}{q - 1}} - 2q \right) t 
\right]
}{
2 \sqrt{2 \pi} \left(-\big((q - 1) q\big)\right)^{3/2}
}.
\end{align}
Using the definitions above, we get the following four saddle point equations for the free energy potential
\begin{align}
\hat{q} &= -2 \times \alpha \left[ \int D t \left( -\Gamma \right) \log \left( 2 + (e^{-\beta} - 1) \left[ \text{erf}\left(\frac{B_{+}}{\sqrt{2}}\right) + \text{erf}\left(\frac{B_{-}}{\sqrt{2}}\right) \right] \right) \right. \notag \\
& \quad + \int D t \left[ \Phi\left(A_{+}\right) + \Phi\left(A_{-}\right) \right] \left\{ \frac{\Psi (e^{-\beta} - 1)}{\left( 2 + (e^{-\beta} - 1)  \left[ \text{erf}\left(\frac{B_{+}}{\sqrt{2}}\right) + \text{erf}\left(\frac{B_{-}}{\sqrt{2}}\right) \right] \right) } \right\} \notag \\
& \quad + \int D t \left( \Gamma \right) \log \left(  2e^{-\beta} - (e^{-\beta} - 1) \left[ \text{erf}\left(\frac{B_{+}}{\sqrt{2}}\right) + \text{erf}\left(\frac{B_{-}}{\sqrt{2}}\right) \right] \right) \notag \\
& \left. \quad + \int D t \left[ \frac{1}{2} \text{erf}\left(\frac{A_{+}}{\sqrt{2}}\right) + \frac{1}{2} \text{erf}\left(\frac{A_{-}}{\sqrt{2}}\right) \right] \left\{ \frac{\Psi (1 - e^{-\beta})}{\left(2e^{-\beta} - (e^{-\beta} - 1) \left[ \text{erf}\left(\frac{B_{+}}{\sqrt{2}}\right) + \text{erf}\left(\frac{B_{-}}{\sqrt{2}}\right) \right] \right)} \right\} \right];
\end{align}

\begin{align}
\hat{R} &= \alpha \left[ \int D t \left( -\Xi \right) \log \left( 2 + (e^{-\beta} - 1) \left[ \text{erf}\left(\frac{B_{+}}{\sqrt{2}}\right) + \text{erf}\left(\frac{B_{-}}{\sqrt{2}}\right) \right] \right) \right. \notag \\
&\left. \quad + \int D t \left( \Xi \right) \log \left(  2e^{-\beta} - (e^{-\beta} - 1) \left[ \text{erf}\left(\frac{B_{+}}{\sqrt{2}}\right) + \text{erf}\left(\frac{B_{-}}{\sqrt{2}}\right) \right] \right) \right]
\end{align}
and
\begin{align}
q &= \int Dz \tanh^2 \left( \sqrt{\hat{q}}z + \hat{R} \right), \\
R &= \int Dz \tanh \left( \sqrt{\hat{q}}z + \hat{R} \right).
\end{align}

\subsection{Bayes-Optimal Case at T=0\label{app:BO}}
\renewcommand\theequation{D\arabic{equation}}
\noindent In the Bayes-Optimal setting, the overlap between the teacher and the student is identical to the overlap between the student of two replicas that is, $q = R$ in the zero temperature limit. We get the following free energy
\begin{align}
& -\mathcal{G} = - \frac{\hat{R}}{2}\left( R + 1\right) + \int Dz\ln2\cosh{(\sqrt{\hat{R}}z + \hat{R})} \notag \\
& + \alpha \times \left[ \int Dt\left[\Phi\left(C_{+}\right)+\Phi\left(C_{-}\right)\right] \log \left(\Phi\left(C_{+}\right)+\Phi\left(C_{-}\right) \right)  \right. \notag \\ 
& \left. +\int Dt\left[\frac{1}{2}\text{erf}\left(\frac{C_{+}}{\sqrt{2}}\right)+\frac{1}{2}\text{erf}\left(\frac{C_{-}^{}}{\sqrt{2}}\right)\right]\log \left(\frac{1}{2} \left[ \text{erf}\left(\frac{C_{+}}{\sqrt{2}}\right) + \text{erf}\left(\frac{C_{-}}{\sqrt{2}}\right) \right] \right) \right],
\end{align}
with 
\begin{equation}
C_{\pm}=\frac{\kappa \pm\sqrt{R}t}{\sqrt{1-R}}.
\end{equation}

\noindent We get the following saddle point equations for the Bayes-Optimal case
\begin{align}
\hat{R} &= 2 \times \alpha \left[ \int D t \left( -\Upsilon \right) \log \left( \Phi\left(C_{+}\right) + \Phi\left(C_{-}\right) \right) \right. \notag \\
& \quad + \int D t \left[ \Phi\left(C_{+}\right) + \Phi\left(C_{-}\right) \right] \left\{ \frac{-\Upsilon}{\Phi\left(C_{+}\right) + \Phi\left(C_{-}\right) } \right\} \notag \\
& \quad + \int D t \left( \Upsilon \right) \log \left(  \frac{1}{2} \left[ \text{erf}\left(\frac{C_{+}}{\sqrt{2}}\right) + \text{erf}\left(\frac{C_{-}}{\sqrt{2}}\right) \right] \right) \notag \\
& \left. \quad + \int D t \left[ \frac{1}{2} \text{erf}\left(\frac{C_{+}}{\sqrt{2}}\right) + \frac{1}{2} \text{erf}\left(\frac{C_{-}}{\sqrt{2}}\right) \right] \left\{ \frac{\Upsilon}{\left(\frac{1}{2} \left[ \text{erf}\left(\frac{C_{+}}{\sqrt{2}}\right) + \text{erf}\left(\frac{C_{-}}{\sqrt{2}}\right) \right] \right)} \right\} \right]
\end{align}
\noindent and 
\begin{align}
R &= \int Dz \tanh^2 \left( \sqrt{\hat{R}}z + \hat{R} \right) ,
\end{align}
where we have introduced the quantity $\Upsilon$ defined as
\begin{equation}
\Upsilon = -\frac{e^{-\frac{(\sqrt{R} \, t + \kappa)^2}{2 (-1 + R)}} \, R 
\left[ 
\left( 1 + e^{\frac{2 \sqrt{R} \, t \, \kappa}{-1 + R}} - 2 R \right) t 
+ \left( 1 + e^{\frac{2 \sqrt{R} \, t \, \kappa}{-1 + R}} \right) \sqrt{R} \, \kappa 
\right]}
{2 \sqrt{2 \pi} \, \left( -((-1 + R) R) \right)^{3/2}}.
\end{equation}
\subsection*{B3. \hspace{0.25cm}Linear Potential\label{app:Vlinear}}
\renewcommand\theequation{E\arabic{equation}}
\noindent For the linear potential, we calculate the below given replicated partition function with potential $V^{(1)}\left(\Delta_{\mu}\right)$ given in \eqref{eq:Vlinear} and $n$ replicas,
\begin{equation}
\mathcal{Z}^n = \exp \left[ 
    -\beta \sum_{a,\mu}^{n,M} V^{(1)}\left(\Delta_{\mu}^a\right) 
\right].
\end{equation}

\noindent We get the following expression for $G_E$ after plugging the potential defined in \eqref{eq:Vlinear} in \eqref{eq:GE_RS_genericpotential}:
\begin{align}
\frac{1}{n}\log G_{E}	&=\int Dt\left[\Phi\left(A_{+}\right)+\Phi\left(A_{-}\right)\right]\log\left\{ \left[\Phi\left(B_{+}\right)+\Phi\left(B_{-}\right)\right]+e^{-\beta \kappa+\frac{\beta^2 (1-q)}{2}}P_1\right\} + \nonumber\\ 
	&+\int Dt\left[\frac{1}{2}\text{erf}\left(\frac{A_{+}}{\sqrt{2}}\right)+\frac{1}{2}\text{erf}\left(\frac{A_{-}^{}}{\sqrt{2}}\right)\right]\log\left\{ e^{\beta \kappa+\frac{\beta^2 (1-q)}{2}}P_2+\frac{1}{2}\text{erf}\left(\frac{B_{+}}{\sqrt{2}}\right)+\frac{1}{2}\text{erf}\left(\frac{B_{-}}{\sqrt{2}}\right)\right\} \nonumber\\
 &+ \text{const}+O(n^2),
\end{align}

\noindent where we have defined $P_1$ and $P_2$ as follows

\begin{align}
    P_1 \, = \frac{1}{2} e^{-\beta \sqrt{q}t}\left[{\rm erf}\left(\frac{B_{-}(\kappa)}{\sqrt{2}}\right) - {\rm erf}\left(\frac{B_{-}(0)}{\sqrt{2}}\right) \right] + \frac{1}{2}e^{\beta \sqrt{q}t}\left[{\rm erf}\left(\frac{B_{+}(0)}{\sqrt{2}}\right) - {\rm erf}\left( \frac{B_{+}(-\kappa)}{\sqrt{2}}\right) \right]
\end{align}

\noindent and 

\begin{align}
    P_2 & \, =  \cosh(\beta \sqrt{q}t)-\frac{1}{2}e^{\beta \sqrt{q}t}{\rm erf}\left(\frac{B_{+}(\kappa)}{\sqrt{2}}\right) + \frac{1}{2}e^{-\beta \sqrt{q}t}{\rm erf}\left(\frac{B_{-} (-\kappa)}{\sqrt{2}}\right) \nonumber \\
    &= \frac{1}{2}e^{\beta \sqrt{q}t}{\rm erfc}\left(\frac{B_{+}(\kappa)}{\sqrt{2}}\right) + \frac{1}{2}e^{-\beta \sqrt{q}t}{\rm erfc}\left(\frac{-B_{-}(-\kappa)}{\sqrt{2}}\right) ,
\end{align}

\noindent using 

\begin{equation}
    B_{\pm}(\kappa) = \frac{\kappa+[\sqrt{q}t \pm \beta (1-q)]}{\sqrt{1-q}}.
\end{equation}
\noindent We now compute the derivative of all terms. We first have that
\begin{equation}
    \partial_q B_{\pm}(\kappa) = \frac{1}{2\sqrt{q}(1-q)^{3/2}}\left[ \mp \sqrt{q}(1-q)\beta + t+\kappa \sqrt{q} \right].
\end{equation}
\noindent Next,
\begin{align}
    \partial_q P_1 =& -\frac{\beta t}{4 \sqrt{q}} e^{-\beta \sqrt{q}t}\left[ {\rm erf}\left(\frac{B_{-}(\kappa)}{\sqrt{2}}\right) - {\rm erf}\left(\frac{B_{-}(0)}{\sqrt{2}}\right)\right] \nonumber \\
    &+ \frac{\beta t}{4 \sqrt{q}} e^{\beta \sqrt{q}t}\left[{\rm erf}\left(\frac{B_{+}(0)}{\sqrt{2}}\right) - {\rm erf}\left( \frac{B_{+}(-\kappa)}{\sqrt{2}}\right) \right] \nonumber \\
    &+ \frac{1}{\sqrt{2 \pi}}e^{-\beta \sqrt{q}t}\left[ e^{\frac{1}{2}B_{-}(\kappa)^2} \partial_q B_{-}(\kappa)- e^{\frac{1}{2}B_{-}(0)^2} \partial_q B_{-}(0)  \right] \nonumber \\
    &+ \frac{1}{\sqrt{2 \pi}}e^{\beta \sqrt{q}t}\left[ e^{\frac{1}{2}B_{+}(0)^2} \partial_q B_{+}(0)- e^{\frac{1}{2}B_{+}(-\kappa)^2} \partial_q B_{+}(-\kappa) \right]
\end{align}
\noindent and
\begin{align}
    \partial_q P_2 =& \frac{\beta t}{4 \sqrt{q}} e^{\beta \sqrt{q}t}{\rm erfc}\left(\frac{B_{+}(\kappa)}{\sqrt{2}}\right) - \frac{\beta t}{4 \sqrt{q}} e^{-\beta \sqrt{q}t} {\rm erfc}\left(\frac{-B_{-}(-\kappa)}{\sqrt{2}}\right) \nonumber \\
    &- \frac{1}{\sqrt{2 \pi}}e^{\beta \sqrt{q}t} e^{\frac{1}{2}B_{+}(\kappa)^2} \partial_q B_{+}(\kappa) + \frac{1}{\sqrt{2 \pi}}e^{-\beta \sqrt{q}t}e^{\frac{1}{2}B_{-}(-\kappa)^2} \partial_q B_{-}(-\kappa).
\end{align}

\noindent Using the above derivatives, we get the saddle point equations below

\begin{align}
\hat{q} &= -2 \times \alpha \left[ \int D t \left( -\Gamma \right) \log \left( \Phi\left(B_{+}\right)+\Phi\left(B_{-}\right)  \ + e^{-\beta \kappa+\frac{\beta^2 (1-q)}{2}} P_1 \right) \right. \notag \\
& \quad + \int D t \left[ \Phi\left(A_{+}\right) + \Phi\left(A_{-}\right) \right] \left\{ \frac{-\Psi + e^{-\beta \kappa+\frac{\beta^2 (1-q)}{2}} P_1 \, (\partial_q P_1 - \frac{P_1 \beta^2}{2})}{ \Phi\left(B_{+}\right)+\Phi\left(B_{-}\right)   \ + e^{-\beta \kappa+\frac{\beta^2 (1-q)}{2}}P_1 }  \right\} \notag \\
& \quad + \int D t \left( \Gamma \right) \log \left(  e^{\beta \kappa+\frac{\beta^2 (1-q)}{2}}P_2+\frac{1}{2}\text{erf}\left(\frac{B_{+}}{\sqrt{2}}\right)+\frac{1}{2}\text{erf}\left(\frac{B_{-}}{\sqrt{2}}\right) \right) \notag \\
& \left. \quad + \int D t \left[ \frac{1}{2} \text{erf}\left(\frac{A_{+}}{\sqrt{2}}\right) + \frac{1}{2} \text{erf}\left(\frac{A_{-}}{\sqrt{2}}\right) \right] \left\{ \frac{\Psi + e^{\beta \kappa+\frac{\beta^2 (1-q)}{2}}(\partial_q P_2 - \frac{P_2 \beta^2}{2})}{ e^{\beta \kappa+\frac{\beta^2 (1-q)}{2}}P_2+\frac{1}{2}\text{erf}\left(\frac{B_{+}}{\sqrt{2}}\right)+\frac{1}{2}\text{erf}\left(\frac{B_{-}}{\sqrt{2}}\right))} \right\} \right];
\end{align}
\begin{align}
\hat{R} &= \alpha \left[ \int D t \left( -\Xi \right) \log \left(  \Phi\left(B_{+}\right)+\Phi\left(B_{-}\right)  \ + e^{-\beta \kappa+\frac{\beta^2 (1-q)}{2}}P_1 \right) \right. \notag \\
&\left. \quad + \int D t \left( \Xi \right) \log \left(  e^{\beta \kappa+\frac{\beta^2 (1-q)}{2}}P_2+\frac{1}{2}\text{erf}\left(\frac{B_{+}}{\sqrt{2}}\right)+\frac{1}{2}\text{erf}\left(\frac{B_{-}}{\sqrt{2}}\right) \right) \right]
\end{align}
and
\begin{align}
q &= \int Dz \tanh^2 \left( \sqrt{\hat{q}}z + \hat{R} \right), \\
R &= \int Dz \tanh \left( \sqrt{\hat{q}}z + \hat{R} \right).
\end{align}

Finally, we get the quenched free energy for the linear potential as

\begin{align}
& -\mathcal{G} = -\text{R}\hat{R} + \frac{\hat{q}}{2}\left( q - 1\right) + \int Dz\ln2\cosh{(\sqrt{\hat{q}}z + \hat{R})} \notag \\
& + \alpha \times \left[ \int Dt\left[\Phi\left(A_{+}\right)+\Phi\left(A_{-}\right)\right]\log \left(  \Phi\left(B_{+}\right)+\Phi\left(B_{-}\right)  \ + e^{-\beta \kappa+\frac{\beta^2 (1-q)}{2}}P_1 \right)\right.\notag \\ 
& \left. +\int Dt\left[\frac{1}{2}\text{erf}\left(\frac{A_{+}}{\sqrt{2}}\right)+\frac{1}{2}\text{erf}\left(\frac{A_{-}^{}}{\sqrt{2}}\right)\right] \log \left(  e^{\beta \kappa+\frac{\beta^2 (1-q)}{2}}P_2+\frac{1}{2}\text{erf}\left(\frac{B_{+}}{\sqrt{2}}\right)+\frac{1}{2}\text{erf}\left(\frac{B_{-}}{\sqrt{2}}\right) \right) \right].
\end{align}